# Highly Ordered Boron Nitride/Epigraphene Epitaxial Films on Silicon Carbide by Lateral Epitaxial Deposition

James Gigliotti[1,2], Xin Li[3,4], Suresh Sundaram[3,4], Dogukan Deniz[1], Vladimir Prudkovskiy[5,1], Jean-Philippe Turmaud[1], Yiran Hu[1], Yue Hu[1], Frédéric Fossard[6], Jean-Sébastien Mérot[6], Annick Loiseau[6], Gilles Patriarche[7], Bokwon Yoon[1], Uzi Landman[1], Abdallah Ougazzaden[3,4] *, Claire Berger[5,1,4] *, Walt A. de Heer[1,8]

[1] School of Physics, Georgia Institute of Technology, Atlanta, Georgia 30332, United States
[2] School of Materials Science and Engineering, Georgia Institute of Technology, Atlanta, Georgia 30332, United States
[3] School of Electrical and Computer Engineering, Georgia Institute of Technology, GT-Lorraine, 57070 Metz, France
[4] Unité Mixte Internationale 2958, CNRS-Georgia Tech, 57070 Metz, France
[5] Institut Néel, CNRS -Université Grenoble Alpes, BP166, 38042 Grenoble Cedex 9, France
[6] Laboratoire d'Etude des Microstructures, ONERA-CNRS, Université Paris Saclay, BP 72, F-92322, Châtillon, France
[7] Centre de Nanosciences et de Nanotechnologies, CNRS, Université Paris-Saclay, route de Nozay, F-91460 Marcoussis, France
[8] Tianjin International Center of Nanoparticles and Nanosystems, Tianjin University, 92 Weijin Road, Nankai District, China

* Corresponding authors:    claire.berger@cnrs.fr; claire.berger@physics.gatech.edu
                            abdallah.ougazzaden@georgiatech-metz.fr

**ABSTRACT:** Realizing high-performance nanoelectronics requires control of materials at the nanoscale. Methods to produce high quality epitaxial graphene (EG) nanostructures on silicon carbide are known. The next step is to grow Van der Waals semiconductors on top of EG nanostructures. Hexagonal boron nitride (h-BN) is a wide bandgap semiconductor with a honeycomb lattice structure that matches that of graphene, making it ideally suited for graphene-based nanoelectronics. Here, we describe the preparation and characterization of multilayer h-BN grown epitaxially on EG using a migration-enhanced metalorganic vapor phase epitaxy process. As a result of the lateral epitaxial deposition (LED) mechanism, the grown h-BN/EG heterostructures have highly ordered epitaxial interfaces, as desired in order to preserve the transport properties of pristine graphene. Atomic scale structural and energetic details of the observed row-by-row, growth mechanism of the 2D epitaxial h-BN film are analyzed through first-principles simulations, demonstrating one-dimensional nucleation-free-energy-barrierless growth. This industrially relevant LED process can be applied to a wide variety of van der Waals materials.

Van der Waals heterostructures have been proposed for many applications, of which, high-performance electronics is the most challenging. Current methods to produce graphene / dielectric heterostructures typically result in contaminated or disordered interfaces. Since epitaxial heterostructures are necessary to enable graphene nanoelectronics, technologically viable production methods are essential.

Epitaxial graphene on silicon carbide (epigraphene, or EG)[1] stands out, because the ultrahigh temperature vacuum growth process (>1500 °C) produces essentially defect-free graphene nanostructures that are crystallographically aligned with the single crystal hexagonal silicon carbide substrate. EG nanostructures have been produced with exceptional room temperature ballistic transport properties, ultrahigh-frequency transistors have been realized, and wafer-scale processing has been demonstrated (ref.[2] and refs therein). That is why epigraphene qualifies as a leading two dimensional (2D) technology platform for future high-performance nanoelectronics.

Further progress requires that Van der Waals semiconductors or dielectrics can be directly grown on EG nanostructures, since all electronic devices require dielectrics for gating, tunneling junctions, and insulation. Common dielectrics, like $Al_2O_3$ and $HfO_2$, use atomic layer deposition (ALD) processes that require chemical modification or roughening of the graphene to facilitate adhesion. These processes cause a variety of problems (see Ref.[3] and refs therein) like charged surface states, surface roughness, impurities and pinholes, that all compromise electronic performance.[3,4]

Here, we concentrate on hexagonal-BN, a semiconductor with a bandgap over 6 eV and a dielectric constant of about 4. It is an isomorph of graphene, chemically inert, that can sustain extremely high temperatures. A lattice mismatch of only 1.7% makes it optimally compatible with epitaxial growth on graphene. Graphene deposited on h-BN, or sandwiched between h-BN crystals shows much improved mobility.[5-10] However these mechanical transfer processes are not scalable, and controlling the graphene-h-BN interface is problematic (trapped impurities, wrinkles, gas filled blisters).[11] Epitaxial h-BN growth has been demonstrated on metal substrates up to wafer scale[12-15] but this process requires transfer onto graphene, resulting in contaminated and otherwise defective interfaces. Since electronic transport in graphene is sensitive to the alignment with h-BN,[16-18] a perfect lattice alignment of the h-BN and graphene is required for reproducible nanoscale electronics applications, therefore transferred materials cannot be used. Early attempts[19-26] to grow h-BN on graphene by conventional chemical phase deposition (CVD) methods or by molecular beam epitaxy[26] were met with limited success in terms of grain size, coverage, graphene integrity or were left with open questions concerning the interface quality and integrity of the graphene layers. Conventional growth of Van der Waals semiconductors ($MoS_2$,[20, 27-28] $WS_2$,[29] $WSe_2$,[20, 30-31]) as well as h-BN[19-25] on graphene often deteriorates from individual single-layer platelets to regions of inhomogeneous thickness as growth proceeds, and, even more seriously results in poor interfaces and even B/N substitution with Carbon of the graphene.

Recently, migration-enhanced metalorganic vapor phase epitaxy (ME-MOVPE), has been used to grow h-BN on silicon[32] and sapphire[33-35] substrates. In this high temperature process, triethylboron (TEB) and ammonia ($NH_3$) are used to produce a multilayer h-BN film on sapphire wafer, informing the present approach.[33] We present here direct evidence that the ME-MOVPE process yields atomically sharp h-BN/graphene epitaxial interfaces, with a templated h-BN growth on graphene. This result, combined with the industrial compatibility and fast growth process, makes MOVPE a method of choice.

Conventional atomic layer depostion is a cyclic processes involving two different precursor molecules in the gas phase that are alternately introduced on a surface, and in which each cycle deposites exactly one atomic layer on top of the film. In contrast, in the migration-enhanced Lateral Epitaxial Deposition (LED) mechanism that we propose here, each cycle "knits" one row of atoms to the edge of the growing multilayer film. The process presented here demonsrtates that the method produces high quality BN-graphene interfaces, and provides a clear path towards further improvement.

**RESULTS AND DISCUSSION**

The LED process proceeds in a MOVPE growth chamber in which the TEB and $NH_3$ volatile molecules, that are the precursors for the h-BN films, are introduced to the extremely hot EG on SiC wafer. Contrary to most CVD processes used to grow h-BN the key steps here are the sequential introduction of the chemicals and the rest time between two injections. A growth cycle consists of first a pulsed injection of TEB that is transported in a flow of hydrogen carrier gas, followed by a purge, followed by a similar sequence for the injection of $NH_3$ (see Methods for details and Ref.33). In this process, the TEB and $NH_3$ precursors alternatively decompose on the hot EG surface, thereby providing active boron then nitrogen atoms to form the h-BN layers.

In the following, we present properties of the h-BN films grown on EG on the (0001) Si-face and the (000$\bar{1}$) C-face of 4H-SiC and on natural EG nanostructures on both the Si- and C-faces. We demonstrate that the h-BN films consists of 2D $sp^2$ layers and that the first h-BN layers are in epitaxy with epigraphene. Figures 1a-d show cross-sectional, high-resolution transmission electron microscopy (HR-TEM) images of the h-BN/EG/SiC heterostructures, composed of h-BN on monolayer EG on the Si-terminated face (Fig. 1a-b) and h-BN on few EG layers on the C-terminated face of hexagonal SiC (Fig. 1c-d). At the interface in particular, the h-BN layers are atomically flat and very well ordered. Close examination of the HR-TEM image of Fig 1d shows AB stacking of the h-BN layers. Fast Fourier Transform (FFT) analysis confirms this stacking (Inset Fig. 1d), and also reveals some admixture of ABC stacking[36] (see also Supplementary Information (SI) Fig. S2 and S3).

This stacking order continues throughout the film and particularly at the critically important h-BN/EG interface (Fig. 1 and Fig. S2 and S3 in SI). This is consistent with the epitaxy of the h-BN film on the graphene (see below). Extended disordered grain boundaries or discontinuities were not observed at the h-BN/graphene interface in any of the twelve 70 nm sections of cross sectional HR-TEM images spanning the 1.5$\mu$m size thin slab that was examined in detail. This suggests that during growth, h-BN sheets lock in place, rather than simply lay on top of the graphene. The local misorientation observed in the uppermost layers in thick h-BN films (Fig. 1a-b), is largely caused by the HR-TEM slab preparation using a focused gallium ion beam, as evidenced by the presence of Ga in those films, (see Fig. S5), due to an insufficient protection by the deposited carbon layer.

Energy Dispersive X-ray Spectroscopy (EDX) in Fig. 1b, identifies chemical components at various depths in the film and correlates with the chemical assignments in Fig. 1a (see Fig. S5 for further analysis). The graphene layers are clearly identified and uniform composition of the h-BN layers is shown.

These results are confirmed by Electron Energy Loss Spectroscopy (EELS), see Fig. S6; in particular, the B-spectra in the film presents a narrow and intense sharp $\pi^*$ peak at 191.1 eV, which is a defining characteristic of $sp^2$-hybridized B atoms. Composition analysis with X-ray photoemission spectroscopy (XPS) has the same line profile as a reference h-BN single crystal (see Fig. S7), and gives the same B/N ratio; the flatness of the h-BN layers also indicates 1/1 boron to nitrogen stoichiometric ratio. XPS profile analysis further indicates a carbon composition of at most about 3% in the h-BN film. This carbon content was determined after etching the spurious surface carbon of h-BN deposited on a sapphire substrate placed next to the EG/SiC substrates during the deposition runs. All the other XPS peaks are similar between h-BN on both substrates, and there is no contribution from $sp^2$ carbon.

The h-BN layer spacing, observed in HR-TEM in Fig. 1a and 1c is 0.345 nm, is very close to the bulk h-BN (0.333 nm), which is consistent with multilayer h-BN on sapphire.[37] On the Si-face EG, the first graphene layer is located 0.43 nm above topmost SiC, in agreement with the spacing reported for quasi-freestanding epigraphene following hydrogen passivation of the SiC surface[38] and lifting up of the buffer layer. Hydrogen intercalation is consistent with the XPS C1s spectra where the characteristic component of the buffer layer is not observed (Fig. S7). It also agrees with LEED observations (Fig. 2b, see below) and with preliminary measurement of a large density of positive charge carriers ($\approx 10^{13}\,cm^{-2}$) for the Si-face EG under h-BN. This indicates that a quasi-freestanding bi-layer graphene is easily produced by the LED method. When helium, instead of hydrogen, is used as the carrier gas, this conversion of the buffer layer to freestanding graphene is expected to not occur, thereby conserving the pristine monolayer graphene on top of the buffer layer.

Figure 2 compiles significant characterization results that are discussed next. (See SI for details). The Scanning Electron Microscopy (SEM) image of Fig. 2a, taken for a 45 nm thick h-BN film grown on Si-face EG, shows a continuous and homogeneous film with a network of three-fold connected pleats. The quasi-hexagonal pleat pattern (see also Fig. S9d) is indicative of an isotropic uniform contraction of 2D layers upon cooling (similarly to multilayers C-face EG

layers$^2$), with no sign of pinning, tearing or large defects. Note that the EG monolayer onto which the h-BN is grown has no pleats, as is common for EG on the Si-face (see Fig. S1a). Low energy electron diffraction (LEED) in Fig. 2b shows the diffraction pattern of the h-BN/EG/SiC heterostructure. The diffraction pattern was recorded with a relatively large electron energy (227 eV), to penetrate through the h-BN and graphene layers, and into the SiC, thereby producing a composite diffraction image. The h-BN diffraction spots are registered with the graphene spots; both are rotated 30° with respect to the SiC. No rings or supplementary spots are observed, that would result if the h-BN layers were rotationally disordered. This observation is consistent with the epitaxy of all three components of the h-BN/EG/SiC heterostructure. The absence of satellites around the graphene spots is consistent with buffer layer lift up by hydrogen intercalation.[39]

High resolution X-ray diffraction (HR-XRD), that quantifies the atomic layer spacing, was performed before and after the h-BN deposition on both C-face and Si-face EG. The h-BN layer spacing was found to be 0.35 nm (Fig. S4), consistent with the HR-TEM (Fig. 1d). The rocking curve measurement of the h-BN (0002) peak (Fig. 2c) gives an extremely narrow full width at half maximum of 0.04° which attests to the h-BN film high degree of parallel order that extends over the width of the X-ray beam (several mm$^2$, that is about the size of the sample).

Raman spectroscopy spectra presented in Fig. 2d were performed before and after h-BN deposition. An example is given in Fig. 2d for a few layers C-face EG. Before h-BN coating, the absence of a Raman D-peak at 1350 cm$^{-1}$ attests to the high quality of the multilayer EG (red trace to be compared with the SiC Raman peaks, blue trace). A large peak develops after h-BN coating at the expected energy of the h-BN E$_{2g}$ peak (1370 cm$^{-1}$). The peak is prominent on areas where the top (disordered) h-BN layers were mechanically removed, attesting to the quality of the h-BN layers closer to the SiC interface. The effect of h-BN deposition on the graphene quality was further investigated by exfoliating the h-BN layer to expose the underlying graphene layer. The Raman spectra show no significant D peak after deposition, indicating that the graphene layer was not damaged in the process (see SI Fig. S12), in agreement with the excellent h-BN/EG interface quality observed in HR-TEM.

The h-BN deposition was repeated on EG samples that were deliberately grown with incomplete graphene layers, to produce EG nanoribbon-like structures on both Si- and C-faces, without resorting to lithographic pattering (Fig. 2c -inset and Fig. S9b & S9c). After processing, the EG structures are covered with a high-quality uniform h-BN layer, while the remaining buffer layer areas are covered with poorly ordered h-BN crumpled sheets. These observations are consistent with the smooth, homogeneous and defect-free EG nanostructures with h-BN nucleation at the edges. In contrast, the buffer layer has an excess of nucleation sites (due to its inherent 6X6 corrugations) that result in the rough h-BN structure. Note that for graphene nanoelectronics a high degree of h-BN order is required only on the EG nanostructures themselves. The growth of smooth layered BN on top of EG is in stark contrast with the preferential h-BN growth on the metallic substrates instead, promoted by the graphene edges in CVD processes (see for instance Refs.40, 41).

The h-BN dielectric quality was tested in a simple 2-probe transport measurement by measuring the I-V characteristic from graphene to a top-gate evaporated on h-BN/EG. The area of the 1 $\mu m^2$ gate is largely compatible with the size of nanoelectronics devices, providing a good indication of the h-BN property in a functional device. The resistance of the h-BN film, measured perpendicularly to the stack, is of the order of 8 GΩ, and breakdown was observed above 165 V (see Fig. S8), which is sufficient to raise the charge density on the EG to about $5 \times 10^{13}$ cm$^{-2}$, *i.e.* an order of a magnitude more than required for intended electronic applications. Nevertheless, larger values are expected with higher quality films. The tested nine top gates gave similar results, showing excellent homogeneity of the dielectric properties, despite the presence of h-BN pleats under some of them (see Fig. S8).

Turning to the growth process, Fig. 3a shows a scanning electron microscope image of the growth on C-face EG after 1200 cycles, showing several merging, quasi-hexagonal h-BN crystallites that nucleated at several random points on the graphene surface. The parallel orientation of the hexagons indicates epitaxy. The size of these crystallites correlates with the number of cycles; in this case each cycle adds on average about one and an half BN rows to the crystallite per cycle.

Growth dynamics on the surface of the EG can therefore be understood in the following way, as depicted in Fig. 3b. At the hot graphene surface where the TEB molecule decomposes after the TEB pulse injection, the liberated highly mobile boron atom migrates over the graphene surface, and attaches to the thermodynamically favorable nitrogen terminated edge of the h-BN film (see theoretical simulation below). To avoid Volmer-Weber 3D growth mode, excess boron atoms have to desorb, which is enabled by the growth-stop step where the chemical residuals (B and methyl compounds) are purged. A $NH_3$ pulse injection follows. The $NH_3$ decomposes on the surface to produce a nitrogen atom that attaches to the boron atom at the edge of the nascent h-BN film. The carrier gas desorbs other volatile decomposition products. In this way, (ideally) a single row of h-BN is added to the edge of the growing film in each cycle of the LED process. This process also applies to subsequent h-BN layer growth, mediated by the fast B and N atoms diffusion on the h-BN surface, similarly to the graphene surface.

We emphasize that the lateral growth process described above is not limited to monolayers but applies to multilayers as well, giving rise to the uniform multilayers observed in Fig. 3a. The process can be understood as follows. Once initiated at a nucleation site, a multilayer growth front will propagate laterally, where B and N atoms alternatively attach to the vertical front. B and N atoms diffusion is mediated as described above, on both on the uncoated graphene and on the growing multilayer h-BN crystal. Hence, the thickness of the film is defined by the properties of the nucleation site as is typical in crystal growth in general. For clean graphene, h-BN nucleation is initiated at the graphene edges and the lithography used to define the graphene nanostructures will need to be optimized to control the BN film thickness.

More systematic studies of different growth conditions and growth rate for fine tuning of the processes will follow in further experiments, in particular by iteratively adjusting the boron precursor flow rate. Systematic optimization for specific applications is routine in all deposition processes. The optimization of the many parameters can be efficiently guided by simulations that provide insight into the processes, as demonstrated below. In particular, the observed hexagonal grains with dendritic edges (Fig. 3a) result from diffusion limited growth conditions, as expected for ordered growth without a free-energy barrier or critical nucleus size, as demonstrated and elaborated by the computer-based simulations below. For Van der Waals material growth in general, grains can vary in shape, from triangular to truncated triangle, to hexagonal depending on the specific growth kinematics, temperature conditions and substrates.[42-43] In the process presented here, the B and N precursors are introduced in the growth chamber, separated with a purge. These growth conditions with optimized precursor ratios allowed to ultimately achieve hexagonal grains with uniform stoichiometry. In addition, the next generation of ME-MOVPE will involve higher temperatures, since an increase in the temperature is known to promote a straight growth front. This is also expected to inhibit the nucleation of subsequent layers and provide a better control of the film thickness.

To explore the structural evolution and energetics of the atomic-scale mechanisms underlying the h-BN film growth processes, we performed first-principles density-functional theory (DFT) simulations; for details see Methods and SI. To this end, we modeled the graphene surface by a periodically replicated supercell (6x12 hexagons comprising 144 C atoms, with an energy optimized interatomic distance $d_{C-C}$ = 1.425 Å) on which we adsorbed a partial strip of h-BN; see the configuration of h-BN/graphene in Fig. 4a. Such initial configuration could be formed by considering a graphene sheet exposing a free edge, serving to initiate the alternating adsorption of B and N rows (following the fast diffusion of the atomic constituents on the bare graphene surface). Our simulations illustrate the atomic-scale mechanisms governing such row-by-row growth process.

To simulate the growth of the h-BN film we focus on two growth modes: (I) where, subsequent to thermal decomposition of the incident (alternating) TEB and $NH_3$ precursors on the graphene T > 1500K heated surface, we consider low fluxes (LF) of B (boron) and N (nitrogen) atoms arriving (through low-barrier thermal diffusion (see SI) on the bare graphene surface) to the vicinity of the h-BN growth front, and (II) where for each of the growth cycles we consider (alternating) high-fluxes (HF) of B and N atoms impinge on the h-BN film growth-front. For our model system each growth cycle entails impingement of 6 B atoms followed by 6 N atoms, resulting, under optimal growth conditions, in the formation of 6 new h-BN hexagons after each cycle. In simulations of the LF mode, we divide (in each cycle) the approach of the 6 B atoms into three stages (two B atoms in each stage, 2B+2B+2B), with the system allowed to structurally and energetically relax to the lowest energy configuration after each stage of 2B atoms diffusion-and-attachment to the film edge. A similar consecutive impingement-followed-by-relaxation procedure is used for the subsequent 6 N (2N+2N+2N) atoms, completing the growth cycle.

LF growth, resulting in perfect hexagonal order at the additional freshly grown h-BN film edge, is shown in the a-b-c sequence in Fig. 4. Note the bonding to graphene of the N atom at the growth edge of the h-BN film, and the debonding (compare Fig. 4b & 4c) of the edge B atoms induced by the impingement of N atoms which bond to the film-edge B atoms; a detailed view of the LF growth mechanism, and corresponding structural parameters are given in Fig. S15. The binding energies, per B atom (5.86 eV) and per N atom (5.50 eV), at the end of the B and N parts of the cycle, respectively, are noted in Fig. 4b & 4c; these binding energies are obtained from the calculated total energy differences corresponding to the system before and after the diffusion and attachment of the graphene-surface-adsorbed B and N atoms to the growth front of the h-BN film (see SI for details). The binding energy per BN unit in the freshly grown h-BN front row is $E_b(BN)$ = 5.86 eV + 5.50 eV = 11.36 eV. Subsequent LF cycles yield a perfectly ordered h-BN film. Interestingly, bondings of B and N atoms is found to occur only at the edge (growth-front) of the h-BN film (see Fig. 4), whereas away from that edge, only weak interaction is found between the h-BN film and the underlying graphene surface (DFT-calculated to be 0.13 eV per BN unit). Furthermore, as observed from comparison of the edge configurations in Fig. 4a & 4b, the attachment of B atoms debonds the edge N atoms from the graphene surface. In turn, we find now graphene-surface-bonded B atoms (Fig. 4b); these atoms debond in the course of the subsequent attachment of N atom as the B & N deposition cycle completes (Fig. 4c). Such bonding-debonding processes at the growth-edge repeat and accompany, throughout, the growth process of the (essentially) perfectly ordered h-BN film. The geometrical parameters corresponding to each of the panels (a-g) are given after Fig. S14.

It is instructive to contrast the above LF results with those obtained from a simulation of the HF growth mode; also termed "inconsecutive" process. Starting from the same initial state we observe at the end of the first (6B+6N) cycle (see Fig. 4a, d, e) a disordered film edge (with a calculated $E_b(BN)$ = 11.15 eV). The subsequent cycle (Fig. 4f & 4g) results in a highly disordered state, with a significantly decreased binding $E_b(BN)$ = 8.51 eV. The superiority of the low-flux (consecutive) growth mode is evident, owing to the structural relaxation/annealing, enabled by the lower rate of impingement of the film's atomic constituents.

We conclude by commenting on the 1D row-by-row growth model[44] of the epitaxial h-BN film, illustrated by our LF simulation. Our simulations uncover ordered growth without a free-energy barrier or critical nucleus size. Indeed, such growth "anomaly" has been long-predicted by classical nucleation theory (CNT)[44, 45] (attributed to J. Williard Gibbs), since in 1D both contributions to the free-energy-change upon phase change – the first contribution associated with *decrease* of the chemical potential caused by formation of a nucleus of the new phase (*e.g.* growth of a crystalline film), and the second one describing the *increase* in energy due to the interface formed between the nucleus and its surrounding – scale linearly with the length of the nucleus (unlike the case in 2D and 3D growth modes), and thus, in 1D, there is no length-scale interplay between the two contributions to the free-energy, and consequently no nucleation barrier, nor a critical-nucleus size, apply here. The above does not preclude the involvement of other activation barriers (kinetic in nature), including, barrier for dissociative adsorption of the B and N precursor molecules, diffusional barriers of the B and N atoms, or chemical attachment barriers. However, the h-BN film growth described in this work appears to not be hampered by such barriers, resulting in an effective growth of a highly-ordered h-BN film on graphene.

**CONCLUSIONS**

In summary, a potentially industrially scalable MOVPE-based lateral epitaxial deposition process has been developed to produce high quality h-BN epitaxial films on EG. The crystalline quality and uniformity of the films and the clean epitaxial h-BN/EG interface were verified using a comprehensive suite of surface probes (SEM, HR-TEM, HR-XRD, XPS, EDX, EELS, LEED, Raman spectroscopy). For short growth time, incomplete coverage reveals that the first h-BN layers have large grains that are rotationally aligned to the SiC substrate. First-principles DFT simulations have uncovered a 1D nucleation-free-energy-barrierless growth mechanism of the epitaxial h-BN film. Atomic-scale details of the row-by-row growth are calculated illustrating the role of atomic-scale relaxation processes at the 2D film growth-front, in correlation with the perfection of structural ordering of the grown h-BN film that indicates the needed control over the fluxes of the constituents (B and N) in the LED process.

The observed uniform thickness and high crystalline quality over large area offer potential for improved reliability and performance in graphene-based nanoelectronics compared to traditional LED methods. Furthermore, since the extent of the prepatterned EG devices are typically less than 100 nm in nanoelectronics, the device edges are ideally suited as nucleation sites to intitiate the line-by-line knitting film-growth, that can be completed in less than 1000 cycles. In an optimized LED process, the cycle time can be optimized further. The h-BN/epigraphene/SiC heterostructure produced here presents the added advantage to be grown on a commercial single crystal semiconductor in an industrial reactor with 3x2-inch wafer capability. Moreover, the processes can be further expanded by growing an epitaxial graphene layer on top of the BN layers, using CVD.[46] In this way graphene/h-BN/graphene heterostructures can be produced. This lateral LED process represents an important enabling step towards the realization of epigraphene nanoelectronics. The technique can be readily adapted to other van der Waals materials, such as $MoS_2$, to enable multilayer Van der Waals epitaxial heterostructures of them as well.

**METHODS**

Epigraphene was produced on commercial chemical mechanical polished (CMP) insulating 4H-SiC substrates (from CREE). Complete few layer graphene films as well as incomplete monolayer graphene domains separated by buffer layer areas were grown on the silicon terminated face (SiC-(0001)), as shown in Fig. S1. Monolayer and multilayer epigraphene domains were produced on the carbon terminated face (SiC-(000$\bar{1}$)) (Fig. S1). The confinement controlled sublimation growth method was used, as described elsewhere,[47] to ensure graphene uniformity across the SiC surface. For this, the vacuum induction furnace was pumped down to $5 \times 10^{-7}$ mbar prior to growth. SiC dies were outgassed at 800 °C prior to ramping to growth temperatures between 1400 °C and 1600 °C for 10-30 min, depending on the desired morphology. Incomplete graphene layers were also grown on 4H-SiC natural steps separated by large atomically flat terraces (5-20 $\mu$m).[48] For comparison, bare CMP 4H-SiC chips were also processed together with the graphene samples in the same MOVPE deposition runs. The above-mentioned graphene morphologies (single and multilayers, partially grown layers and nanostructures) were used for h-BN film-growth.

Migration enhanced metal organic vapor phase deposition process was applied for h-BN growth using an Aixtron close coupled showerhead 3x2" MOVPE reactor. Prior to growth, the chamber was heated to 1270°C in hydrogen environment at 85 mbar. A triethylboron (TEB) preflow (5-10 s) preceded the alternate introduction of TEB and ammonia ($NH_3$) to nucleate and grow h-BN layers. A cycle consists of first a pulsed injection of TEB (1-3 s), where the total carrier gas flow rate was maintained at 20 SLPM with a TEB flow rate of 60 $\mu$mol/min. This is followed by a 1-3 s purge of the volatile components. Then $NH_3$ is introduced for 3-6 s (flow rate: 1.35 SLPM, carrier gas: 20 SLPM), followed by a 1-3 s purge similarly to a standard ALD process. The cycle is repeated for full coverage. h-BN films were grown with thicknesses ranging from about 1 to 50 nm. The growth rate in the c-axis direction was 5 nm/h as determined by AFM (scratch profile). The growth rate on EG was about 1/3 of that observed on sapphire (15 nm/h),[33] likely due to a lower nucleation density on the graphene surface migration enhanced MOVPE process is applied to reduce excess nucleation on the surface of growing BN layers. Experience from BAlN and BGaN nitride MOVPE deposition studies indicate that simultaneous introduction of precursors leads to lower nitride quality achieved than achieved with the alternation protocol.[49]

For the cross-sectional HR-TEM study, 70-80 nm thick lamella were prepared by FIB milling following deposition of 50 nm of amorphous carbon (Fig. 1a-b, S2, S5, S6) or Pt (Fig. 1c-d, S3) to protect the h-BN and graphene from ion bombardment and to reduce delamination. The images were taken along the $[11\bar{2}0]$ zone axis on a Titan Themis microscope at 200 kV. EDX was performed to determine the extent of Ga-ion damage, differentiate graphene and h-BN, and to examine the chemical uniformity of the film.

The first-principles electronic structure calculations were performed with the Vienna *Ab initio* Simulation Package (VASP). A plane-wave basis with a kinetic energy cutoff of 400 eV, projector augmented wave (PAW) pseudopotentials method[50] with the PBE generalized gradient approximation (GGA) for the exchange–correlation potential, and including van-der Waals interactions, were used.[51] We employed a lattice parameter of 2.468 Å (C-C bond length of 1.425 Å) for the two-dimensional unit cell of a graphene's hexagonal lattice; these values were found from our DFT energy optimization (minimization) calculations. A unit cell of the graphene lattice contains two carbon atoms. We employed a graphene calculational unit cell comprised of 6×6 hexagons (72 C atoms) for the boron and nitrogen atom adsorption calculations (see SI), and a 6×12 (144 C atoms) graphene calculational cell for the h-BN film-growth simulations; the calculational cell was repeated periodically. An initial h-BN strip of 6×6 hexagons was constructed from 42 boron and 42 nitrogen atoms with the same lattice parameter as graphene, and it was positioned on the 6x12 graphene surface, and subsequently energy optimized; the initial BN strip (see Fig. 4a) comprised 912 valence electrons. Most of our calculations were performed using a single k-point sampling (that is Γ-point calculations); we checked that the results of h-BN growth remained essentially the same, by employing (3×3×1) and (6×6×1) sampling of the surface Brillouin zone (see SI).

In simulation of the h-BN film growth process, for both the low-flux (consecutive) and high-flux (inconsecutive) modes, the adsorbed B or N atoms were initially located (adsorbed) on the bare graphene surface at a distance up to 5 Å from any atom of the h-BN film growth-front, to ensure no direct initial bonding to the film. The attachment of the atoms to the film-growth front occurred in the course of (unconstrained) DFT energy minimization of the total system.

## ACKNOWLEDGMENTS

Financial support was provided by the AFOSR under grant #FA9550-13-0217 and NSF #1506006. The work of UL and BY was supported by the AFOSR grant # FA9550-15-1-0519. Calculations were carried out at the GATECH Center for Computational Materials Science. This work was also made possible by the French American Cultural Exchange council through a Partner University Fund project. CB, VP, FF. and AL acknowledge funding from the European Union's Horizon 2020 research and innovation program under grant agreements No. 696656 (Graphene Core 1) and No 785219 (Graphene Core 2).

**SUPPORTING INFORMATION**

Additional data on (1) epigraphene morphology, (2) h-BN crystal structure, (3) h-BN chemical and dielectric characterization, (4) h-BN surface morphology, (5) epigraphene characterization, (6) theory.

The authors declare no competing financial interest.

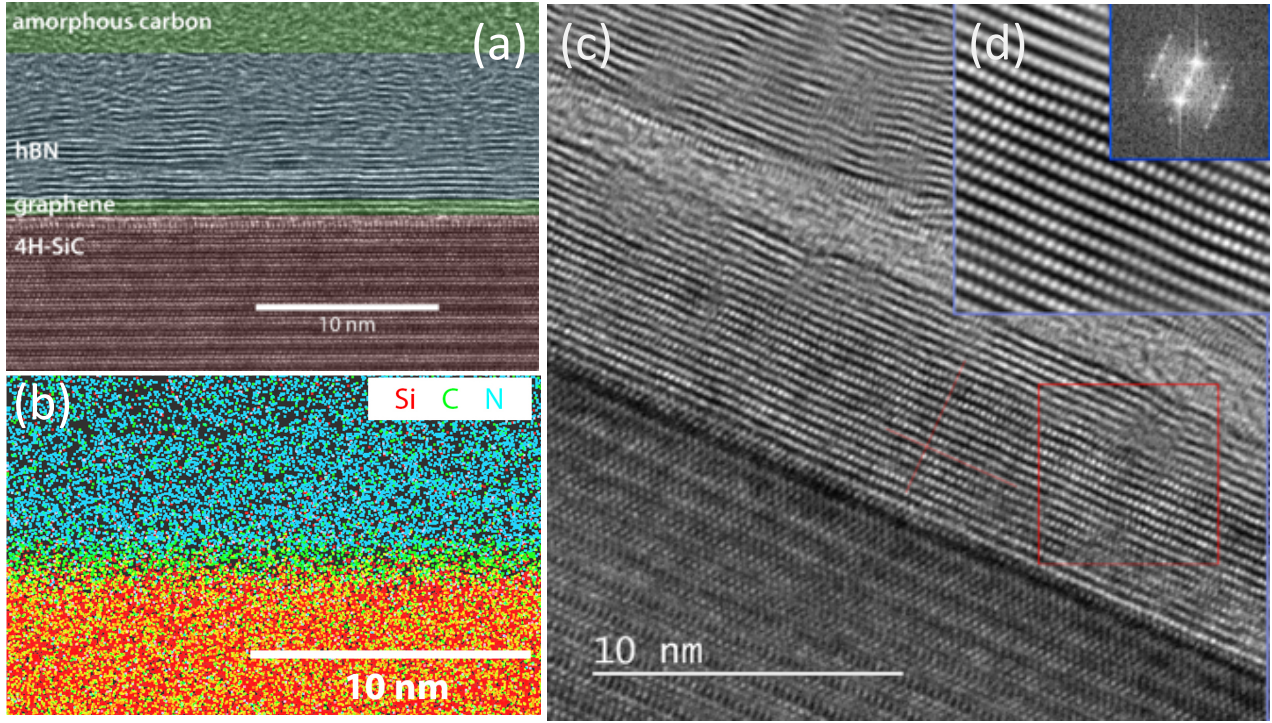

**Figure 1:** *Cross-sectional HR-TEM of h-BN on epigraphene along the zone axis*. **(a)** Colored high resolution image depicting coherence in the h-BN layer grown on one (possibly 2 layers) EG on the Si-face. Identification of the layers is obtained from the EDX analysis in (b). Disorder of the top h-BN layers is mainly due Ga ion damage from the FIB preparation. The spacing of the first graphene layer is 0.43 nm, indicating hydrogen intercalation at the SiC/EG interface. **(b)** EDX analysis of the HR-TEM image in (a) chemically resolving the N atoms in the h-BN layer, the C- in the graphene layer and the Si-atoms in the SiC layer (sensitivity-limited detected B is shown in Fig.S5, as well as FIB induced Ga contamination). **(c)** HR-TEM of another heterostructure, confirming that the h-BN layers are well ordered. **(d)** Zoom in the region in the red square in (c), showing an interlayer spacing of the h-BN layer of 0.345 nm and dominant AB stacking. Inset: Fast Fourier Transform showing AB stacking with some admixture of ABC stacking. (The disordered area between the well-ordered layers is a separation between layers due to delamination during the thin slab preparation.)

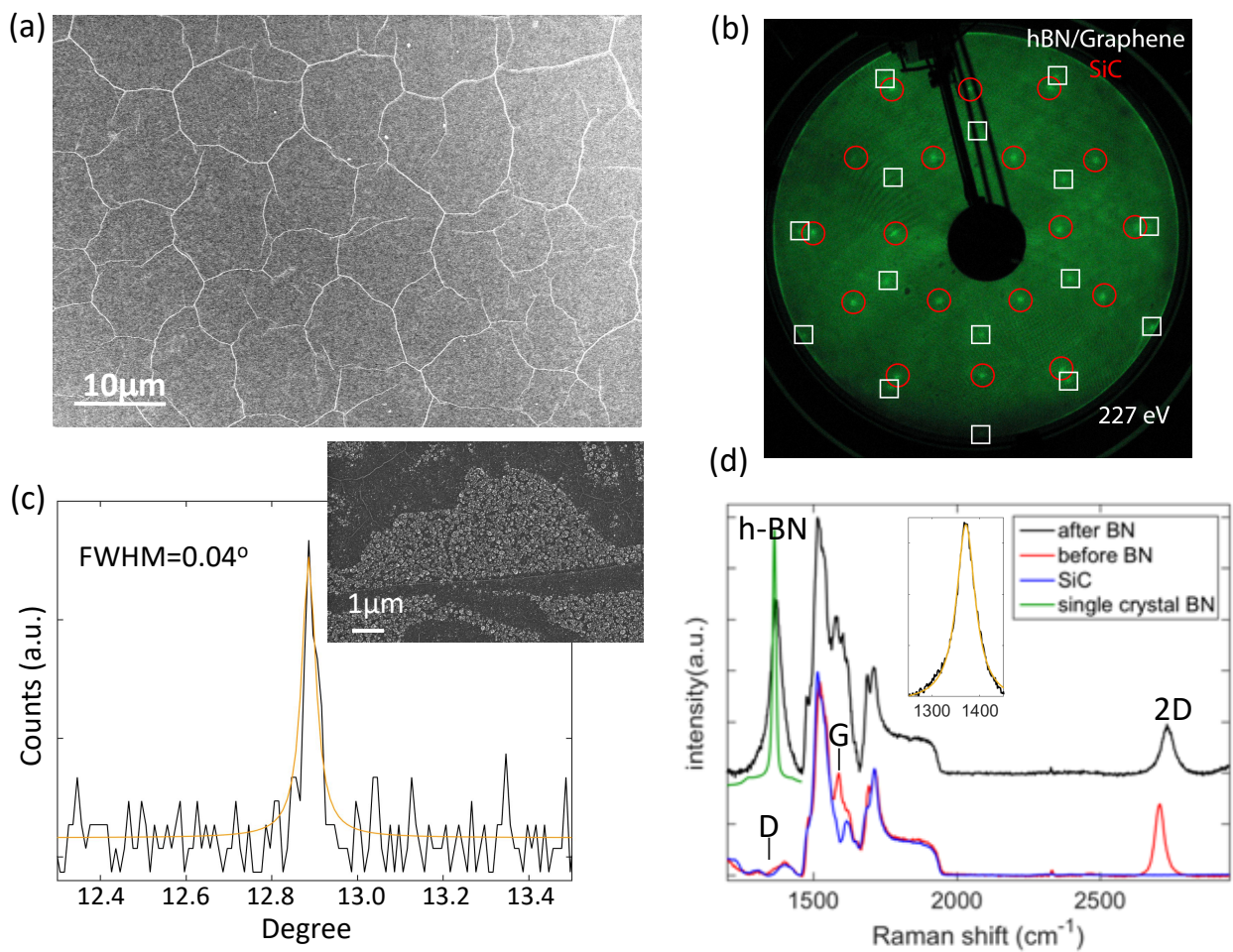

**Figure 2:** *Analysis of a h-BN/EG/SiC heterostructure*. **(a)** SEM image of an h-BN film prepared on a monolayer epigraphene of the Si-face (SiC(0001)), showing a homogeneous full coverage. White lines are pleats. **(b)** Low energy electron diffraction performed with a 227eV electron beam, showing the expected h-BN, EG and SiC diffraction spots. The relative azimuthal orientation of the spots and the absence of rings and other anomalies confirm the epitaxy of the heterostructure h-BN/EG/SiC. **(c)** The HR-XRD rocking curve has a full width at half maximum (fwhm) of 0.04° demonstrating the extreme flatness of the film layers. Inset: SEM image of h-BN coating on a partially graphitized SiC surface, composed of EG ribbons (grown on natural SiC substrate step walls) separated by buffer layer regions, showing the smooth h-BN/EG surface (dark areas), compared with the rough h-BN/buffer layer surface caused by the very high density of nucleation sites on the buffer layer. **(d)** Raman spectra of few graphene layers on the carbon face (red) compared to the bare SiC spectrum (blue). The Raman spectrum after h-BN growth (black) displays a large peak at 1370 cm$^{-1}$ that is ascribed to the h-BN $E_{2g}$ due to the absence of a D peak in the as-grown EG. (For this spectrum, the disordered top h-BN layers are mechanically removed) Inset: Lorentizan fit of the peak at 1370 cm$^{-1}$ (FWHM=47.3 cm$^{-1}$).

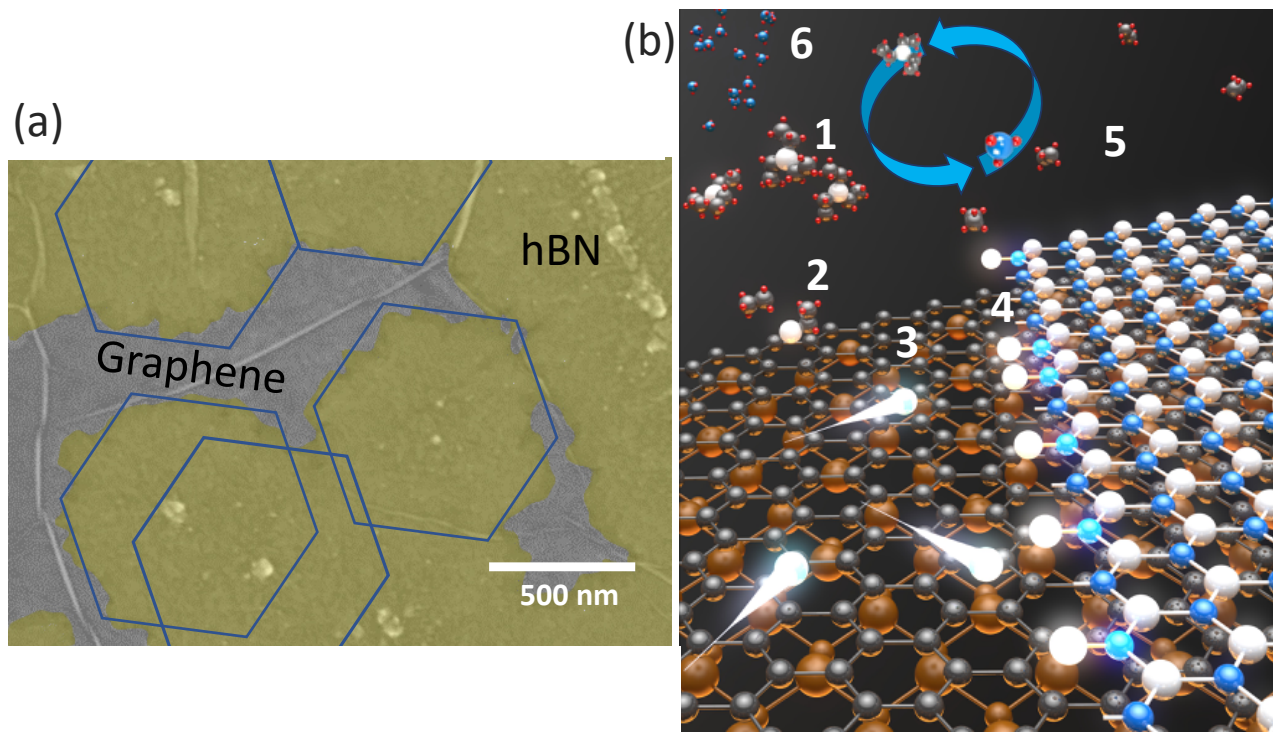

**Figure 3: (a)** Colored SEM image showing a partially grown multilayer h-BN film on a multilayer C-face graphene after 1200 LED cycles. The size of the hexagonal 2D h-BN crystallites corresponds with the expected size for roughly one atomic front growth per cycle. The hexagons merge (they do not overlap) ultimately producing a single continuous sheet (see text). The large pleats on the bare graphene areas are commonly seen for the graphene on the C-face. **(b)** Schematic of the formation of the epitaxial 2D heterostructure using the lateral ALD process. The process cyclically injects TEB then ammonia. In the depicted TEB phase of the cycle, TEB (1) decomposes on the hot epigraphene surface liberating ethane (2) that escapes. The free, highly mobile boron atom (3) skids on the epigraphene to an awaiting nitrogen atom at the edge of the growing film (4). The process terminates when all nitrogen sites are passivated, and is ready for the second half of the cycle where the remaining TEB and ethane are purged (5) and ammonia is introduced (6) to add a fresh row of nitrogen to the boron saturated edge by the analogous decomposition process.

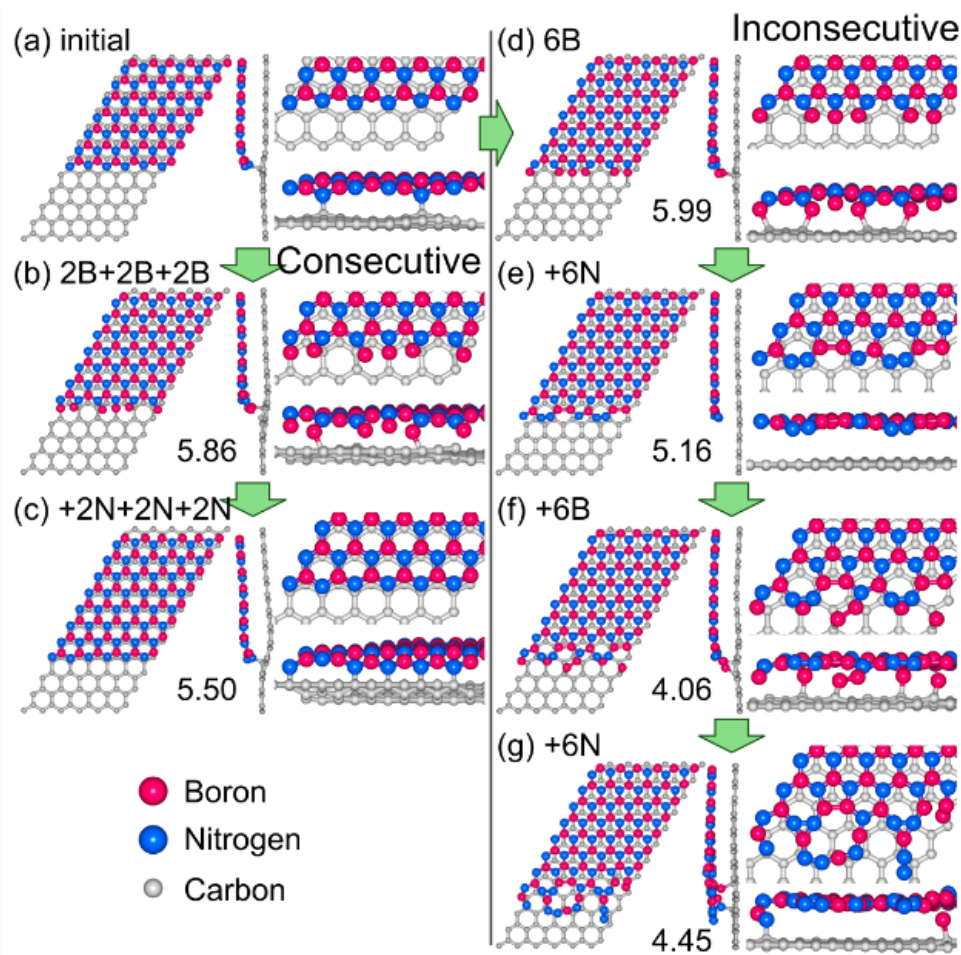

**Figure 4:** First-principle calculated mechanism of the growth of a h-BN film on epigraphene. The DFT model system consists of a graphene surface comprised of 6x12 hexagons, periodically replicated in the xy plane, partially covered by an h-BN strip. Boron atoms are depicted as red balls, nitrogen atoms in blue, and carbon atoms in grey color. In each of the panels (except for panel a) we include the DFT-calculated binding energy (per B or N atom) in eV units, associated with the corresponding stage of h-BN film growth shown.

**(a)** The energy optimal initial configuration, showing a partially grown h-BN film adsorbed on the graphene surface. A top view is shown on the left. In the middle, we show a side view illustrating the 'floating' nature of the partially grown h-BN film. At the top right a zoomed view (from above) of the edge of the h-BN film is depicted, showing nitrogen termination of the growing front, with two of the N atoms bonded to the graphene surface. Below the above view we show an enlarged view taken normal to the growing film edge, illustrating the weak coupling between the h-BN film and the graphene substrate.

In **(b)** we depict the optimal configurations resulting from exposure of the initial configuration (in a) to a low flux (LF) of B atoms. The small flux is modeled *via* the consecutive adsorption of boron atoms (2B atoms at each adsorption stage), with the system allowed to relaxed to the energy-optimal configuration after each adsorption step (for a detailed view of the system after each of the three consecutive B-adsorption steps see Fig. S15). The configuration shown in (b) is the one obtained at the end of the consecutive adsorption 2B+2B+2B process (resulting in 6 adsorbed B atoms corresponding to the 6 edge-exposed N atoms). Note that in this configuration all the adsorbed B atoms are bonded to both the nearest edge-N atoms of the film, as well as anchored (to various degrees) to the underlying graphene surface.

**(c)** The consecutive adsorption of nitrogen atoms (low flux) results in the formation of an added row of perfect h-BN to the growing film. Five out of six of the N atoms at the bottom edge of the freshly grown hexagon row are anchored to the graphene surface.

**(d-g)** An inconsecutive growth process, corresponding to high flux, HF, of arriving boron atoms followed by HF of impinging nitrogen atoms; that is, in each B/N cycle all the boron, as well as all the subsequent nitrogen atoms, arrive (essentially) simultaneously to the growing edge of the h-BN film (with no, or incomplete at best, intervening structural relaxations taking place at the growth front, see initial configuration in a followed by the configurations in d), resulting in defective growth of the film (see configuration e). The subsequent HF adsorption cycle yield a disordered film edge (see g).

Supporting Information

# Highly Ordered Boron Nitride/Epigraphene Epitaxial Films on Silicon Carbide by Lateral Epitaxial Deposition

*James Gigliotti, Xin Li, Suresh Sundaram, Dogukan Deniz, Vladimir Prudkovskiy, Jean-Philippe Turmaud Yiran Hu, Yue Hu, Frédéric Fossard, Jean-Sébastien Mérot, Annick Loiseau, Gilles Patriarche, Bokwon Yoon, Uzi Landman, Abdallah Ougazzaden, Claire Berger, Walt de Heer*

**I. Epigraphene Morphology**

The morphology and electronic properties of epigraphene[1] depend on its growth conditions and growth face, as shown in Fig. S1. On the silicon-terminated face of 4H-SiC (that is (0001)), the buffer layer is the first graphene-like layer at the interface with SiC. It exhibits strong interactions to the SiC substrate, which induces structural corrugation and semiconductor properties. The subsequent layer has the iconic graphene band structure. Graphene on the Si-face is flat and shows no pleats. It covers uniformly the SiC wafer and traces of bilayer are observed, at steps edges. If graphitization is stopped before graphene grows onto the SiC terraces, graphene nanoribbons can be formed on sidewall of SiC steps, separated by buffer layer on the terraces. On the carbon-terminated face (SiC($000\bar{1}$)), multilayer graphene (with rotational layer stacking) is produced, yielding a pleated structure due to the negative in-plane coefficient of thermal expansion and weak interlayer forces.[1]

**II. h-BN Crystal Structure**

Cross sectional HR-TEM images in Fig. S2 and S3 show a layered h-BN structure with a pristine graphene-h-BN interface and no visible grain boundaries. Fast Fourier Transform (FFT) analysis confirms BN layer stacking. The atomic positions can be seen within the graphene and h-BN layers in Fig. S3b, confirming epitaxy between the h-BN, graphene, and SiC.

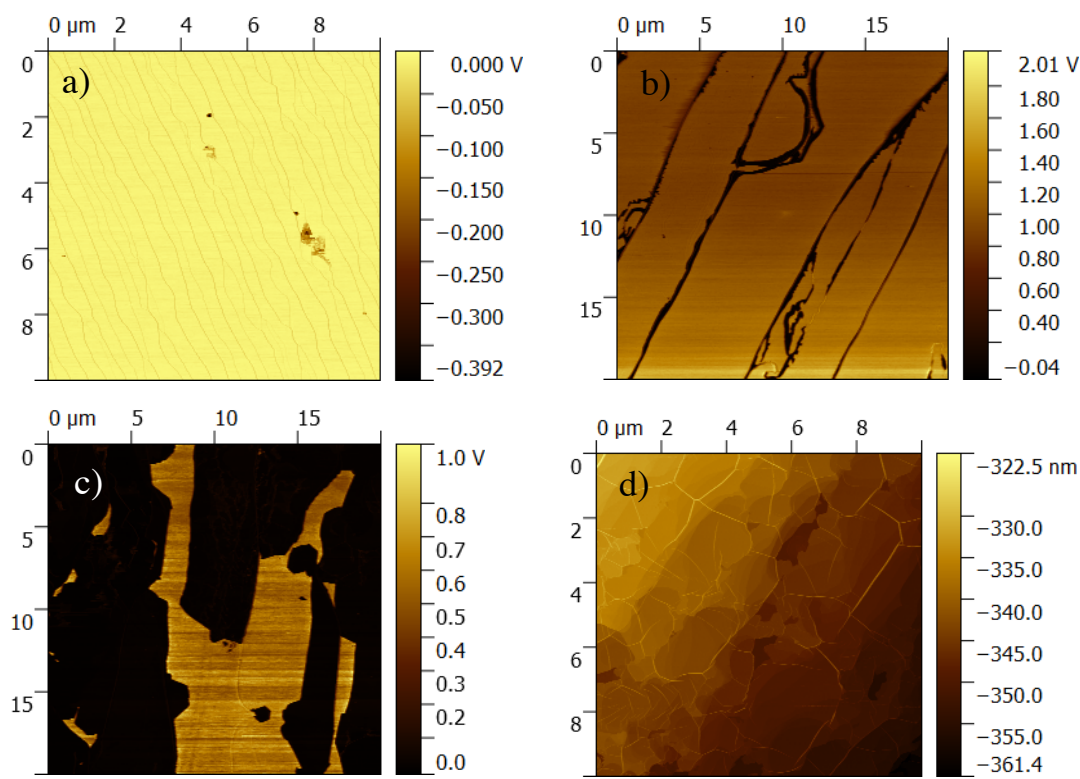

**Fig. S1:** Representative atomic force microscopy images: (a-c) lateral force microscopy (LFM) and (d) topography of epigraphene. In the case of LFM, regions with tip-surface interactions near zero are graphene while the regions of higher tip interaction are buffer layer or SiC. Note that the contrast in the images was chosen to highlight the graphene features and is not consistent. (a) Monolayer graphene grown on the Si-face of SiC, exhibiting continuous graphene (bright contrast). The faint lines are SiC steps that graphene covers. No pleats are observed. (b) Graphene nanoribbons (dark lines) grown on natural SiC step edges on the Si-face. The terraces are covered with the buffer layer (c) Monolayer graphene domains (black) on the carbon-face of SiC. The light areas are bare SiC, since the graphene grows directly on the C-face, with no buffer layer. (d) Multilayer epigraphene on the carbon face, showing that graphene drapes over the entire surface with a pleated surface morphology.

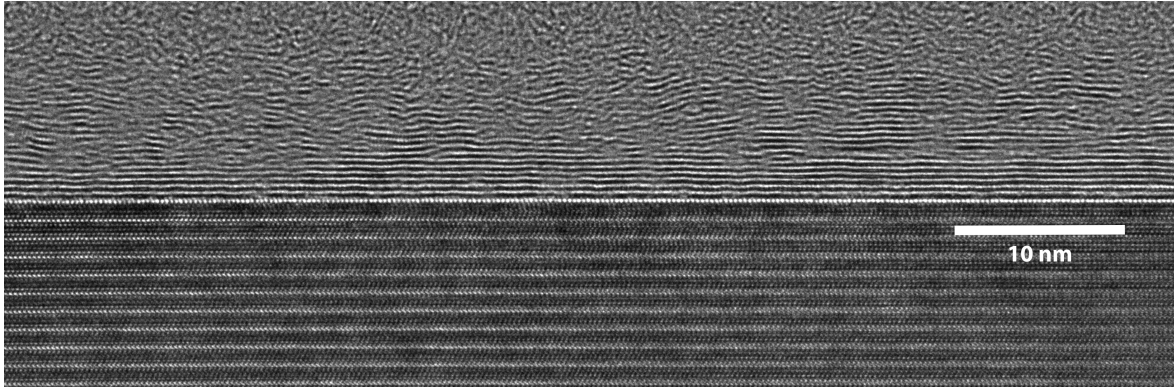

**Fig. S2:** Representative HR-TEM cross-section of h-BN on monolayer graphene on 4H-SiC. Disorder in the top layers is largely induced by the FIB preparation in this case (carbon protection coating – see Fig. S5).

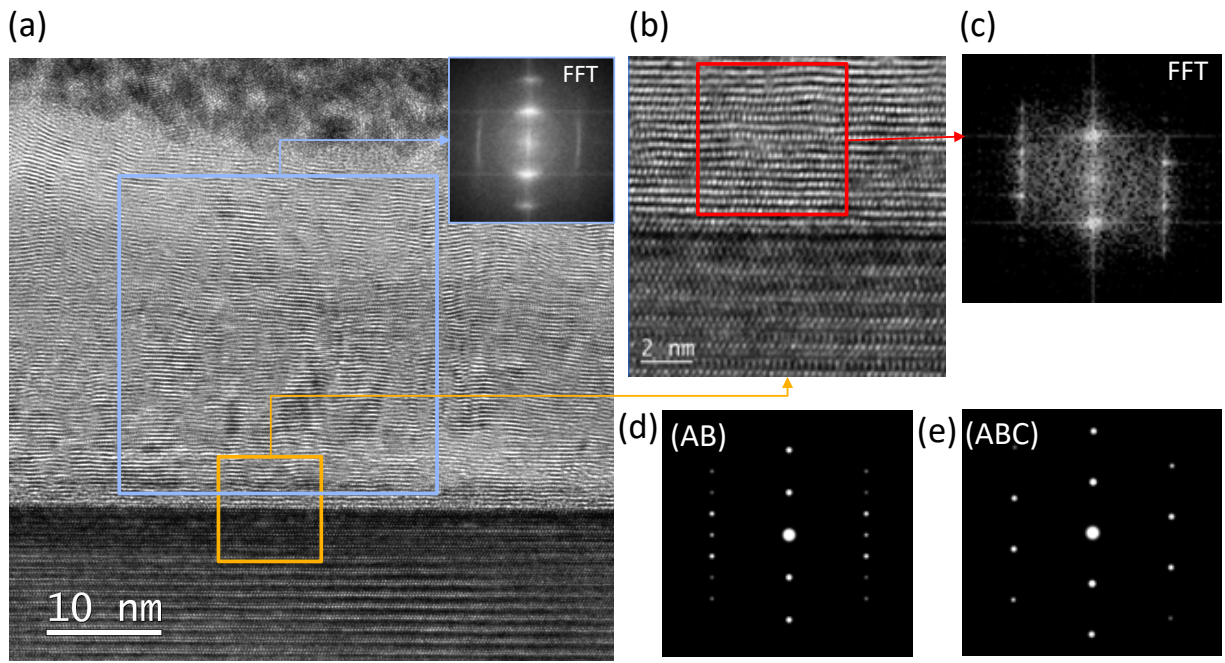

**Fig. S3:** Cross-sectional HR-TEM of a thick h-BN film on epigraphene prepared by FIB using a Pt coating (a) Large scale image showing the whole h-BN/EG/SiC heterostructure. Inset FFT of the h-BN region boxed in blue (b) zoom on the region boxed in yellow in (a), showing atomic resolution and the parallel layers close to the interface. (c) FFT of the region boxed in red close to the interface, showing h-BN (AB) stacking with some admixture of r-BN (ABC). (d)-(e) numerical simulation of diffraction for (d) h-BN (AB) stacking and (e) r-BN (ABC) stacking.

High resolution X-ray diffraction (HR-XRD) scans covering the entire sample (about 15 nm h-BN on a few layer graphene on the C-face), as shown in Fig. S4, confirm the highly ordered h-BN layers. Scans were completed in triple-axis configuration with monochromatic Cu-Kα radiation on a Panalytical X'pert Pro MRD. The h-BN (0002) peak is located at 2θ = 25.35 ° corresponding to an interlayer spacing of 3.5 Å. This represents a 4.8% expansion over bulk h-BN (3.33 Å), and agrees with the HR-TEM results. The interplanar spacing of graphene is nearly identical to h-BN. In order to confirm that the peaks come from the h-BN film, background HR-XRD scans were taken prior to h-BN growth, and no peak is observed for monolayer graphene. Graphene peaks are only observed for few layer graphene due to the small graphene cross section. HR-XRD spectra acquired from granular h-BN films showed a significantly reduced h-BN peak, confirming that the pleated morphology observed in Fig. 2a is a characteristic of h-BN layers.

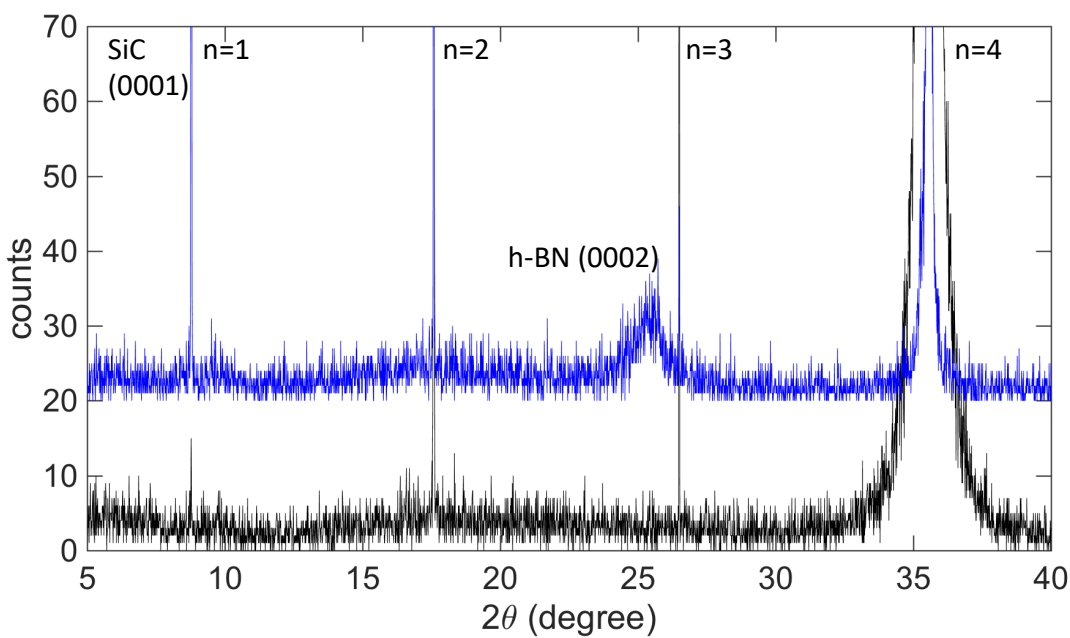

**Fig. S4:** Representative HR-XRD taken from a 15 nm thick h-BN film deposited on few layer graphene grown on the C-face. The h-BN (0002) peak is located at 2θ=25.5°. The sharp, quasi-forbidden SiC peaks (SiC(0001) and its multiple orders) were used to ensure alignment to the substrate.

## III. h-BN Chemical and Dielectric Characterization

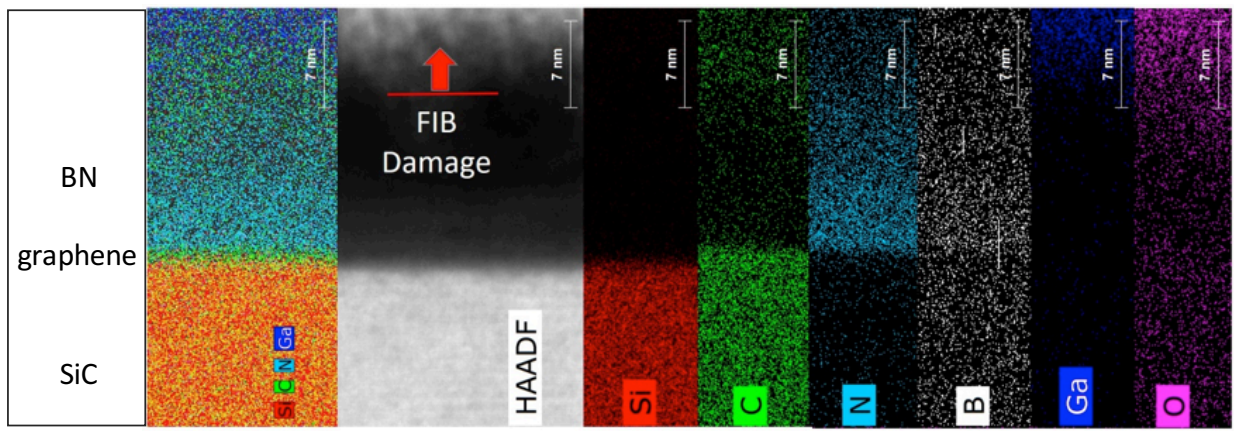

**Fig. S5**: EDX spectra taken during cross sectional TEM reveal sharp interfaces between the SiC, graphene, and h-BN. As can be seen, Boron does not provide sufficient contrast against the background signals due to its low atomic number. The gallium, carbon, and oxygen incorporation into the top, distorted h-BN layers was caused by the FIB preparation (carbon protection coating, same slab as in Fig. 1a-b main text).

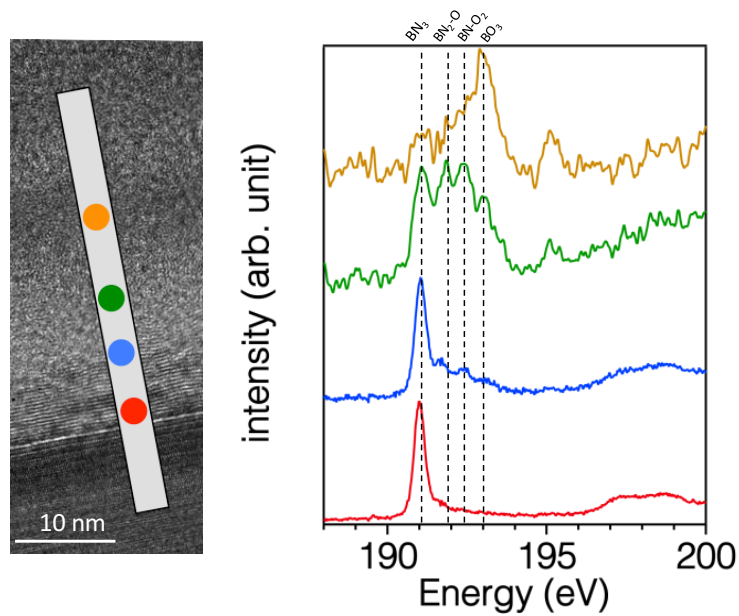

**Fig. S6**: Electron Energy Loss Spectroscopy (EELS) on the same thin h-BN/EG/SiC slab as Fig. S5, and Fig. 1a-b confirms the EDX results, and the $sp^2$ character of the BN film. The red spectrum close to the EG interface presents an intense and sharp $\pi^*$ peak at 191.1 eV that is a fingerprint of $sp^2$-hybridized B atoms in the hexagonal BN network, (*i.e.*, a trigonal $BN_3$ bonding environment). The blue spectrum (7 nm above the SiC) and green spectrum (12 nm above the SiC) show the emergence of three additional features related to defective boron nitride film with some oxygen substitution to nitrogen. The yellow spectrum (20 nm above the SiC) presents the signature of a heavy $BO_3$ contamination [2-3], due to thin slab processing.

Energy dispersive X-ray spectroscopy (EDX) and Electron Energy Loss Spectroscopy (EELS) were performed on cross sectional samples (same as Fig. 1a-b, main text), as shown in Fig. S5 and S6. A sharp interface between the SiC, graphene, and h-BN is observed. The cross-section was produced by a gallium ion beam (FIB) that penetrated through an amorphous carbon protection layer, contributing to the layer distortion. Thus, high gallium and carbon and oxygen are incorporated in the top h-BN layers.

Fig. S7 shows selected XPS spectra taken from a 10 nm h-BN film on a Si-face EG after a mild Ar sputtering (standard cleaning procedure) on a Thermo K-alpha XPS. Once the adventitious surface carbon is removed, h-BN, graphene, and SiC XPS signatures are identified. The B1s and N1s peaks each present one significant component, centered at 190.8 eV and 398.5 eV, respectively, corresponding to literature values for $sp^2$ hybridized BN. [4-6] Satellite peaks, occurring at 198 eV and 406 eV are present for both B and N, respectively, due to the π- π* transition that are not present in cubic ($sp^3$) BN. [7] The peaks do not shift throughout the thickness of the film indicating homogeneous chemical structure. Comparing the normalized areas with an h-BN single crystal yields the same B/N concentration, indicating 1:1 stoichiometry, in accordance with the flat layers observed in HR-TEM and HR-XRD. The C1s spectrum reveals two dominant components: graphene at 284.5 eV and SiC at 283.3 eV. Components of the buffer layer[8] could not be identified, likely due to intercalation that lifts the buffer layer. The bulk of the h-BN film contains at most about 3 at% carbon, most probably due to retained carbon from the triethylboron ligands, as determined by examining h-BN films grown on sapphire. The Si 2p spectrum shows only a pure SiC contribution at 100.8 eV. The oxygen spectrum (not shown) is primarily composed of a small concentration of metal oxides at 532.3 eV. Granular h-BN films grown on the buffer layer and on SiC were also analyzed and reveal $sp^2$ bonding (see Fig. S11), supporting the crumpled sheet morphology (see Fig. S9).

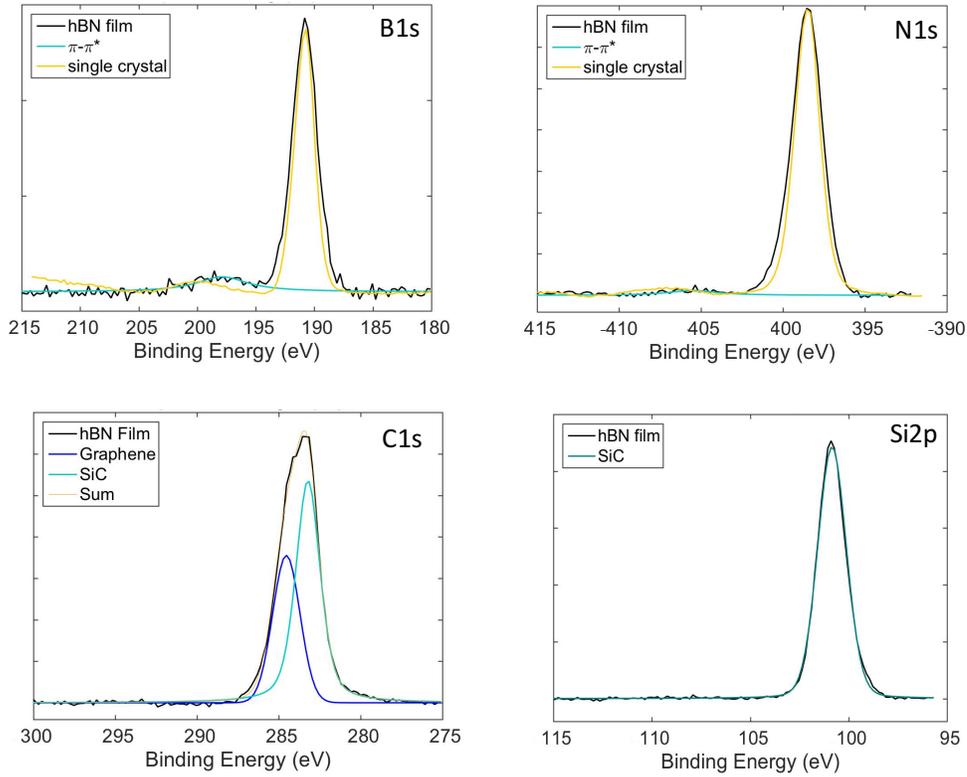

**Fig. S7:** XPS spectra taken after Ar sputtering on a 10nm h-BN grown on monolayer graphene on the Si-face. The B and N spectra have the same profile as h-BN single crystal, confirming the 1:1 stoichiometry of the h-BN film. The π-π* satellite peaks confirm sp$^2$ bonding. The h-BN single layer crystal peaks were shifted by about 1eV to match the film peaks; the shift is likely due to different experimental conditions. Graphene and SiC are identified in the carbon spectrum.

The dielectric quality of the h-BN film sandwiched between EG and a 1 $\mu$m x 1 $\mu$m top gate was tested in a 2-probe current – bias voltage measurement, as described in the main text. The gates were produced by e-beam evaporation of Cr/Au (5 nm / 30 nm) onto lithographically patterned PMMA/MMA and lift-off, in the same geometry as an EG/dielectric/metal top-gated field effect transistor. A Pt coated AFM tip in contact mode was use to contact the metal gate. The resistance of the tip contact to Au was first measured by making contact on a Cr/Au pad prepared on the graphene (Resistance < 5 k$\Omega$). Fig. S8 shows I-V characteristics for a 45nm thick h-BN film (thickness measured by AFM at the step of a h-BN scratch). Similar I-Vs are obtained for all 9 gates shown in the AFM image in the inset.

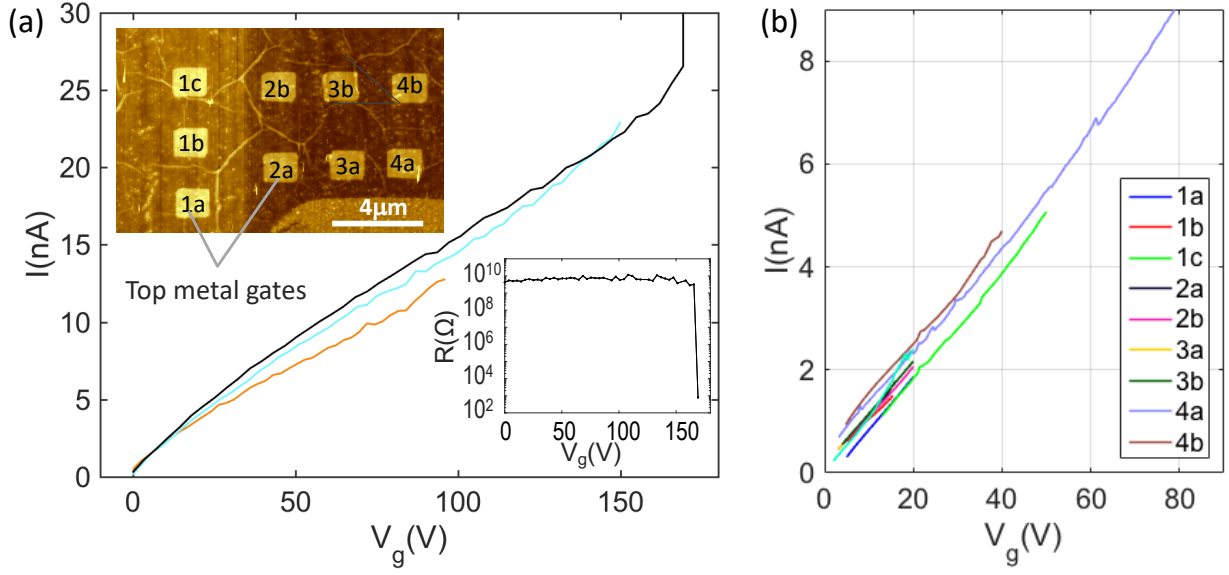

**Fig. S8:** Current-Voltage characteristic of the h-BN film on epigraphene, measured from the top metal gate evaporated to top of h-BN to epigraphene below h-BN. (a) Each trace is a sweep at a higher voltage $V_g$ and shows excellent reproducibility until the breakdown voltage of 165V. (bottom inset) The gate resistance R=dV/dI=8 GΩ is roughly constant up to the breakdown. (top inset) AFM topographic image of the measured device showing the nine 1μm×1μm evaporated metal dots on top of h-BN/EG. Traces in (a) are for device 3b. (b) I-V characteristics for all the dots, showing consistently about the same R=$V_g$/I≈10 GΩ resistance.

## IV. h-BN Surface Morphology

Continuous h-BN films with pleated surface morphology grow on graphitized regions of both the SiC silicon and carbon faces, as shown in Fig. S9, and exhibit no dependence on the number of graphene layers. Like graphene, h-BN has a negative in-plane coefficient of thermal expansion that induces a large compressive strain upon cooling from the high temperature of h-BN growth. The weak interlayer interaction allows the sheets to slide with respect to each other to accommodate this strain without cracking, forming pleats. The pleats often radiate with tri-fold symmetry (see Fig. S9d) from sites where the h-BN is pinned, indicative of biaxial in-plane stress,[9] like for multilayer graphene.[1] The height of the pleats is 5-15nm as determined by AFM, shown in Fig. S10, in agreement with previous studies.[10] The 1260°C growth temperature of h-BN does not enable silicon sublimation from the SiC substrate, and therefore precludes additional graphene growth as the cause of the pleating. Note also that pleats are not observed on graphene grown on the Si-face before h-BN coating (see Fig. S1).

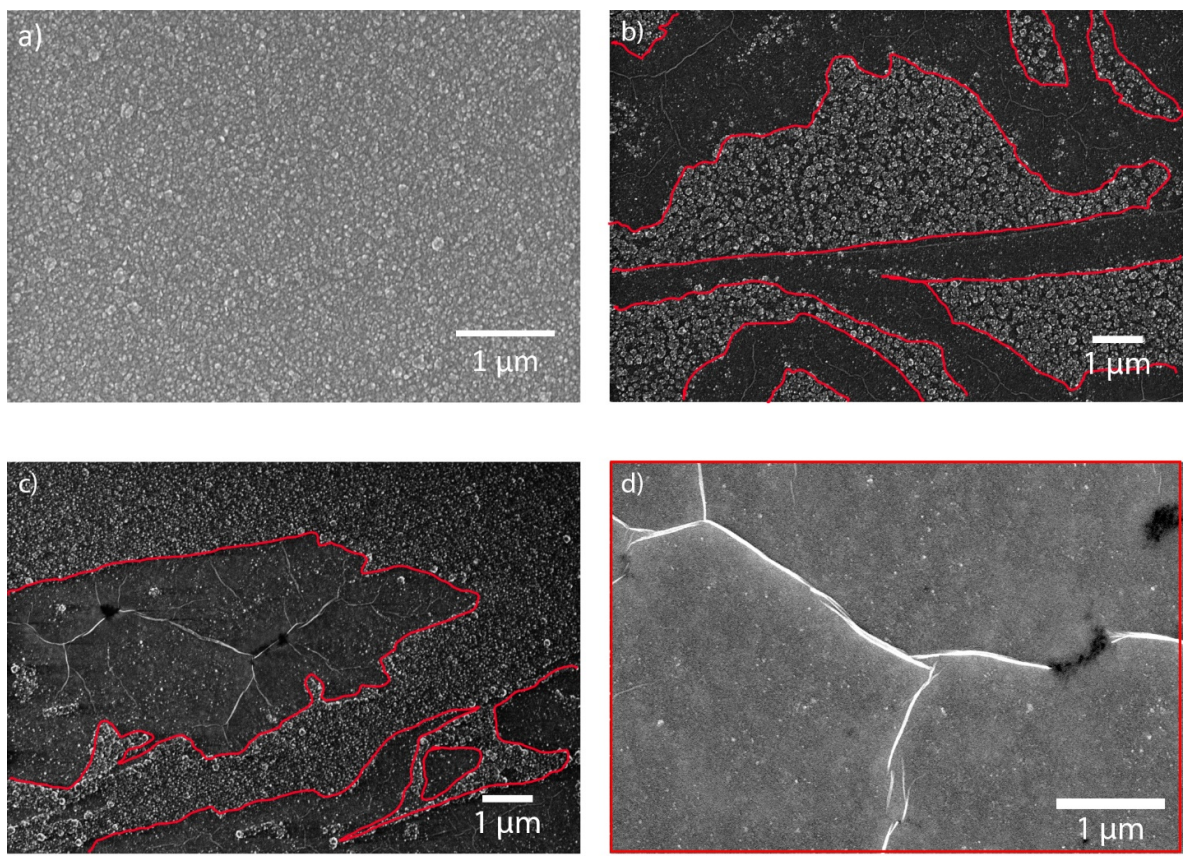

**Fig. S9:** Scanning electron images of the surface morphology of h-BN deposited on partially graphitized SiC. (a) h-BN on bare SiC, (b) h-BN on partially grown graphene structures on Si-face natural (disordered) steps (graphene is the smooth darker areas, outlined in red, the rest is the buffer layer), (c) h-BN on partially grown monolayer graphene on the C-face (graphene is the smooth darker areas, outlined in red, the rest is SiC), and (d) h-BN on uniformly covered monolayer Si-face graphene. The h-BN exhibits selective growth with a flat pleated surface morphology on graphitized regions and a granular structure on the SiC and buffer layer. The granular structure on the buffer layer is larger than on SiC.

The flat and pleated h-BN films are predominantly observed on EG domains. This is in stark contrast to other typical film deposition processes that do not nucleate or grow uniform films on graphene (see for instance[11]). In fact, the inert surface of graphene is so unfavorable to film growth in general and presents so few nucleation sites that technologies have been developed using graphene as a mask for dielectric coatings.[11-12] In order to grow uniform Van der Waals materials, film growth must occur under conditions which enable sufficient adatom mobility to encourage lateral growth of each layer and platelet shifting to accommodate misalignments between impinging grains. The high temperature and carrier flow used here, in addition to eliminating vapor phase reactions by separating the precursors, contribute to low growth rate conditions (through low precursor flux and high adatom mobility), which favors selective lateral growth on the flat EG and h-BN $sp^2$ surfaces (see theory section, main text).

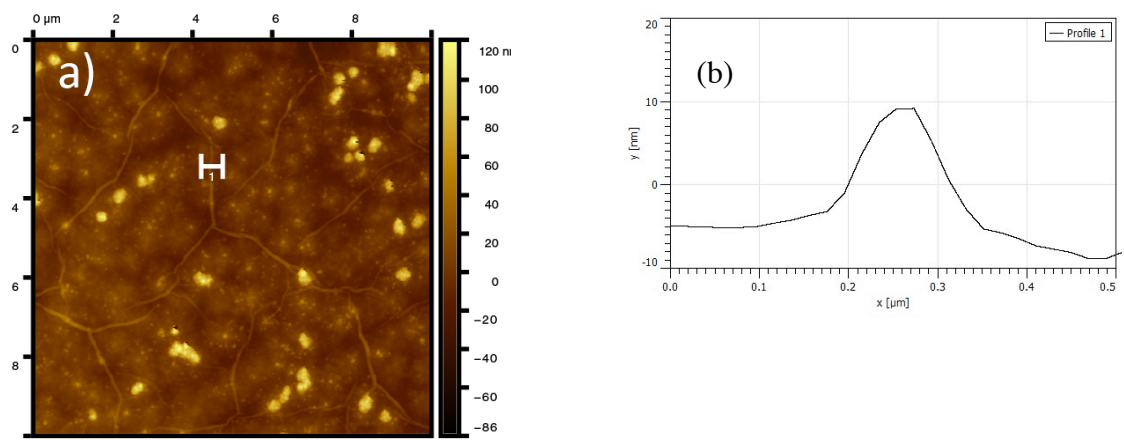

**Fig. S10:** (a) AFM topography 10 $\mu$m x 10 $\mu$m scan showing pleated h-BN morphology on monolayer Si-face graphene (25 nm thick h-BN). (b) profile along the white horizontal line in (a) across a pleat. The h-BN pleat heights are 5-15 nm.

Nanostructured graphene exhibits nearly pristine h-BN surfaces (Fig. S9b, c), in contrast with the bare SiC (Fig. S9a, S9c) and the buffer layer regions (Fig. S9b), where instead of the pleated morphology, a granular structure is observed. The h-BN clusters are actually crumpled $sp^2$ h-BN sheets, as no $sp^3$ hybridized BN is identified in XPS, even on highly granular surfaces (see Fig. S11). The h-BN crumpling is associated to a larger SiC lattice mismatch compared to graphene, or to higher nucleation density due to reduced adatom mobility, as expected from the corrugated surface (1 Å) of the buffer layer and the $sp^3$ hybridization of SiC. Further experiments conducted with low temperature TEB preflow have shown high particle densities on all surfaces, supporting that the high temperature mobility is imperative for h-BN growth. The h-BN clusters also often decorate SiC step edges in thin h-BN layers (Fig. S9b, S9c), but were not observed at the graphene interface in HR-TEM cross sections confirming that they occur only on the h-BN top surface; crumpling at the graphene edges could be due to incomplete h-BN layers rearrangement during cooling. Note that, as flat h-BN growth is favored on the graphene nanostructures themselves, this will facilitate implementation into the EG nano-sized devices required for graphene-based electronics.

For thicker h-BN films (>20nm or so), clusters do not decorate step edges, as shown in Fig. S10, but the concentration and size of crumpled granules on the h-BN pleated regions increase. Stacking faults, as observed in HR-TEM (Fig. S3a), induce stress within the h-BN film that may cause the h-BN sheets to crumple. For these thick h-BN films, Raman spectra of as-deposited films show a broad band instead of the large peak shown in Fig 2d, which may be related to the disordered top layers.

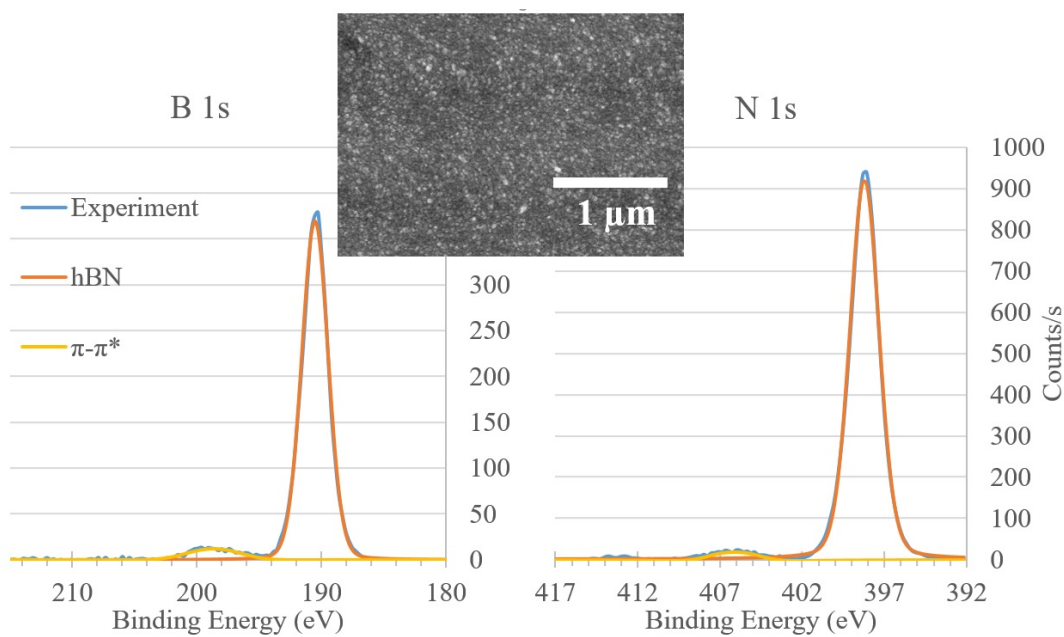

**Fig. S11:** XPS taken from the granular h-BN surface shown in the SEM image in the inset indicating $sp^2$ bonding, even in this case.

## V. Epigraphene Characterization

Raman spectroscopy is widely used to quantify disorder in graphene. To determine the damage caused by processing, a monolayer graphene on the Si face was measured before and after the h-BN deposition as follows. After the deposition, the h-BN layer was mechanically exfoliated to expose the underlying graphene layer so that its Raman spectrum can be measured. As shown in Fig. S12, the graphene Raman spectrum (laser excitation 532 nm) shows no D peak (at 1350 cm$^{-1}$), confirming high graphene quality. The D-peak does not increase after deposition, indicating that the graphene layer was not damaged by the process. Significantly, the 2D peak (at 2715 cm$^{-1}$) is slightly red-shifted (2717 to 2714 cm$^{-1}$) and broadened from (46 to 57 cm$^{-1}$), consistent with the conversion of the buffer layer into a quasi-freestanding graphene layer (see main text) by hydrogen passivation of the SiC surface.

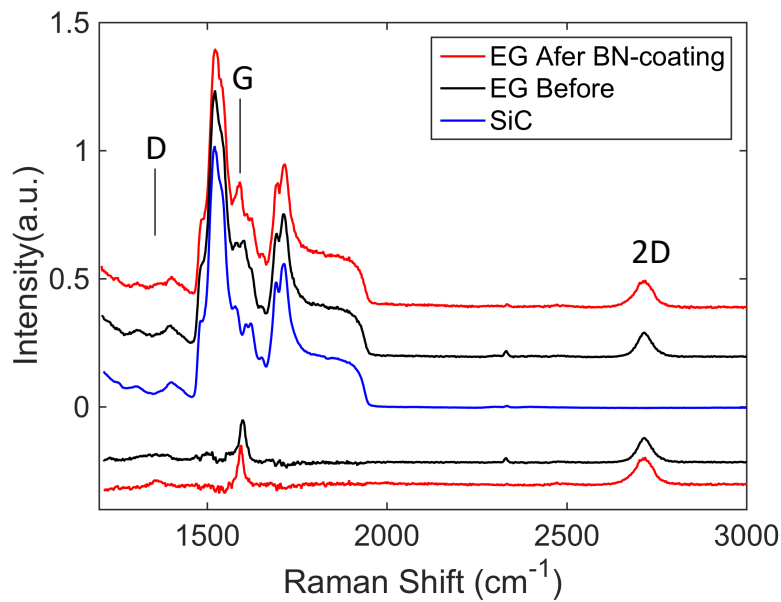

**Fig. S12**: (three upmost spectra) Raman spectra of the graphene before (red) and after (black) h-BN growth on a monolayer graphene on the Si-face, in comparison to bare SiC (blue). (bottom most pair) Same as above, but with the SiC spectra subtracted. The absence of a D peak, within the precision of the subtraction, clearly shows that the LED process did not damage the graphene. For the red spectra, the h-BN layer was removed by mechanical exfoliation to expose the graphene.

## VI. Theory

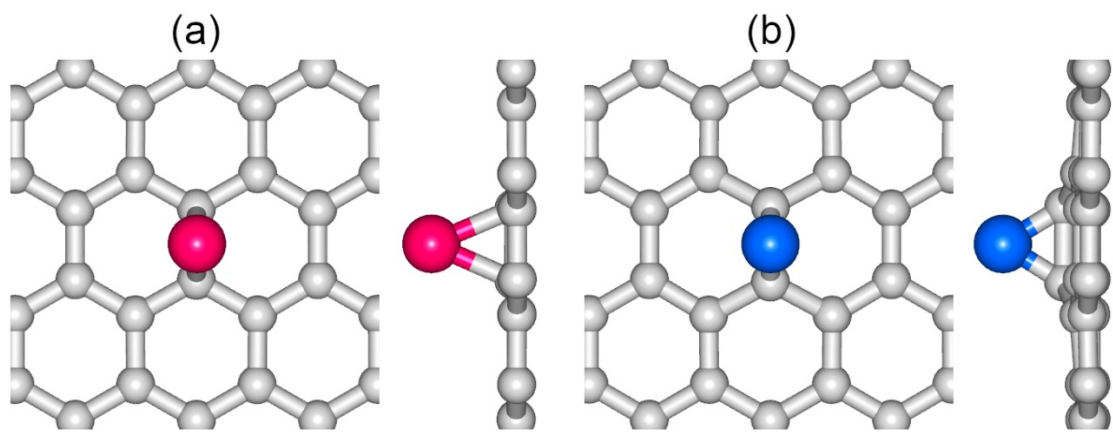

**Fig. S13:** Adsorption configuration of: (a) a Boron (red) and (b) a Nitrogen (blue) atom on the graphene lattice. In each case both a top-down and side-view are shown. Away from the adsorbed atom, $d_{C-C}$ (graphene) =1.425Å

The geometry in (a) $d_{B-C} = 1.833$ Å; $d'_{C-C}$ (under the B atom) = 1.462 Å

The geometry in (b) $d_{N-C} = 1.458$ Å; $d'_{C-C}$ (under the N atom) = 1.537 Å

Away from the adsorbed atom, $d_{C-C}$ (graphene) = 1.425 Å

The adsorption energy of B atom on the graphene(6x12) is 1.67 eV (gamma-point only)

The energy barrier of B atom diffusion on the graphene is 0.06 eV (gamma-point only)

The adsorption energy of N atom on the graphene(6x12) is 4.13 eV (gamma-point only)

The energy barrier of N atom diffusion on the graphene is 0.97 eV (gamma-point only)

The adsorption energy of B atom on the graphene(6x12) is 1.16 eV (6x6x1 k-points)

The energy barrier of B atom diffusion on the graphene is 0.10 eV (6x6x1 k-points)

The adsorption energy of N atom on the graphene(6x12) is 3.64 eV (6x6x1 k-points)

The energy barrier of N atom diffusion on the graphene is 0.75 eV (6x6x1 k-points)

**Evaluation of binding and adsorption energies**

The adsorption energy of n B atoms on the graphene area of the g/h-BN system is given by :

$$E_{ad}(g/hBN + nB) = E(g/hBN) + nE(B) - E(g/hBN + nB), \qquad (1)$$

where $E(g/hBN)$ is the energy of the g/h-BN system- that is, partial h-BN strip adsorbed on the graphene (g) surface and $E(B)$ is the energy of a gas-phase B atom, and $E(g/hBN + nB)$ is the energy of the g/h-BN system with n B atoms adsorbed on the graphene area.

For the B atoms that are adsorbed on the graphene surface (away from the partial h-BN strip) we can approximate $E_{ad}(g/hBN + nB)$ as $E_{ad}(g + nB)$, that is we can take

$E_{ad}(g/hBN + nB) \cong E_{ad}(g + nB)$, where $E_{ad}(g + nB)$ is calculated as follows:

$E_{ad}(g + nB) = E(g) + nE(B) - E(g + nB)$. As a result:

$$E_{ad}(g + nB) \cong E(g/hBN) + nE(B) - E(g/hBN + nB). \qquad (2)$$

In the following we will write $E_{ad}(g + nB)$ as $E_{ad}(nB])$ for simplicity

$$E_{ad}(nB) \cong E(g/hBN) + nE(B) - E(g/hBN + nB). \qquad (3)$$

The energy of the g/h-BN system with n B atoms adsorbed on the graphene area is then

$$E(g/hBN + nB) \cong E(g/hBN) + nE(B) - E_{ad}(nB) \qquad (4)$$

The binding energy of n B atoms on the h-BN edge of the system is

$$E_b(nB) = E(g/hBN + nB) - E(g/hBN/nB) \qquad (5)$$

Using Eq. (4), Eq. (5) becomes

$$E_b(nB) \cong \left[E\left(\tfrac{g}{hBN}\right) + nE(B) - E_{ad}(nB)\right] - E(g/hBN/nB), \qquad (6)$$

with $E_{ad}(nB) = nE_{ad}(B)$,

$$E_b(nB) \cong \left[E\left(\tfrac{g}{hBN}\right) + nE(B) - nE_{ad}(B)\right] - E(g/hBN/nB). \qquad (7)$$

For n N atoms the formula for the binding energy of n N atoms to the h-BN growing edge is:

$$E_b(nN) \cong \left[E\left(\tfrac{g}{hBN}\right) + nE(N) - nE_{ad}(N)\right] - E(g/hBN/nN). \qquad (8)$$

In our calculation, $E(B)$, $E(N)$, $E_{ad}(B)$, and $E_{ad}(N)$ are -0.25eV, -0.35 eV, 1.67 eV, and 4.13eV, respectively.

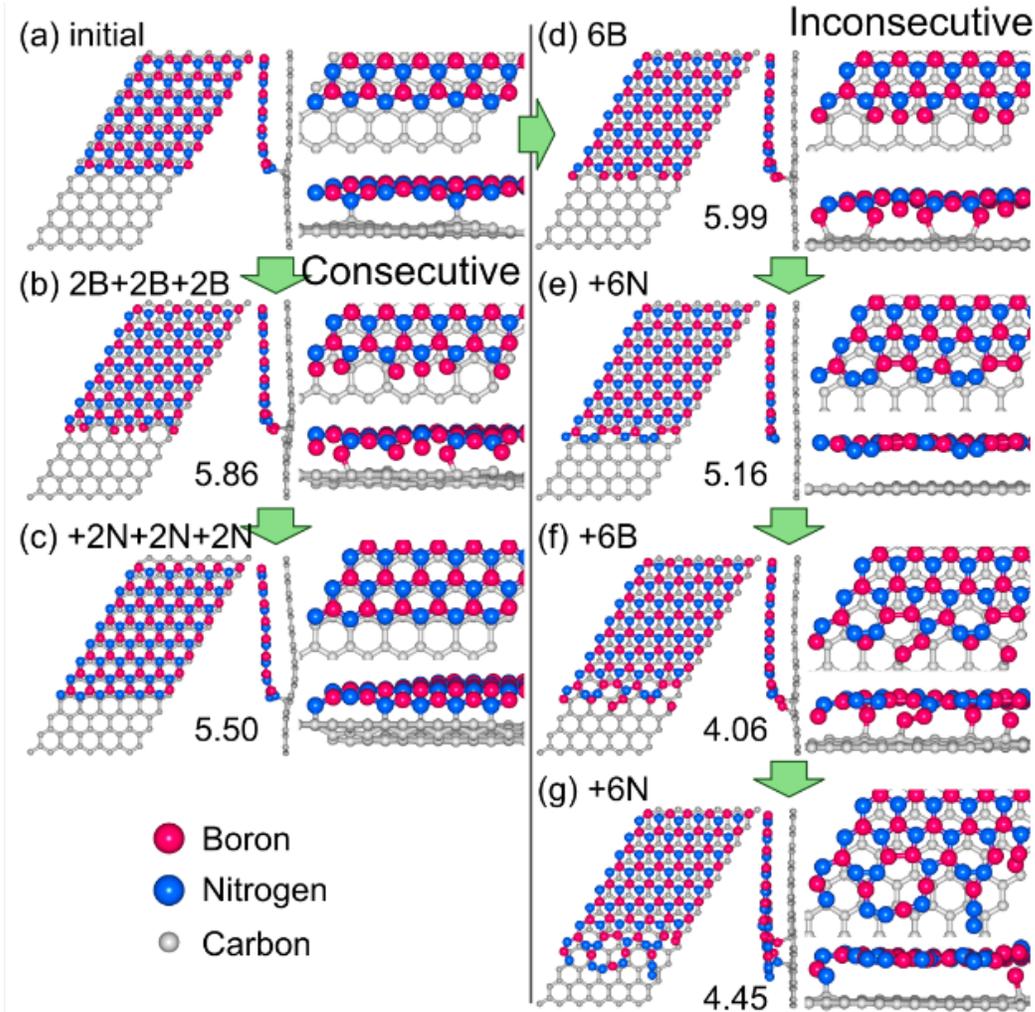

**Fig. S14:** Geometrical structures and energies for one-dimensional h-BN growth in low-flux (consecutive) and high-flux (inconsecutive) growth modes. For the reader's convenience we include here Fig. 4 of the main text. (see detailed caption in the main text).

Geometrical parameters:

(a) In the middle of the graphene layer (3$^{rd}$ row from the top, from left to right, in Å ):

$d_{B-C}$ = {3.457, 3.454, 3.456, 3.453, 3.453, 3.453}, <$d_{B-C}$> = 3.454

$d_{N-C}$ = {3.367, 3.373, 3.366, 3.379, 3.366, 3,373}, <$d_{N-C}$> = 3.371

$d_{B-N}$ = {1.426, 1.428, 1.426, 1.428, 1.426, 1.428}, <$d_{B-N}$> = 1.427

At the bottom edge (from left to right, in Å ):

$d_{B-C}$ = {2.637, 2.601, 3.106, 2.637, 2.601, 3.106}, <$d_{B-C}$> = 2.781

$d_{N-C}$ = {2.523, 1.552, 2.532, 2.522, 1.552, 2.533}, <$d_{N-C}$> = 2.202

$d_{B-N}$ = {1.402, 1.402, 1.427, 1.402, 1.402, 1.427}, <$d_{B-N}$> = 1.410

(b) At the bottom edge (from left to right, in Å ):

    $d_{B-C}$ = {2.538, 1.681, 1.834, 2.538, 1.681, 1.834}, <$d_{B-C}$> = 2.018

    $d_{N-C}$ = {3.605, 2.657, 2.823, 3.605, 2.657, 2.823}, <$d_{N-C}$> = 3.028

    $d_{B-N}$ = {1.436, 1.435, 1.489, 1.436, 1.435, 1.489}, <$d_{B-N}$> = 1.453

(c) At the bottom edge (from left to right, in Å ):

    $d_{B-C}$ = {2.616, 2.608, 2.474, 2.522, 2.525, 2.471}, <$d_{B-C}$> = 2.536

    $d_{N-C}$ = {1.539, 2.568, 1.538, 1.551, 1.558, 1.551}, <$d_{N-C}$> = 1.717

    $d_{B-N}$ = {1.397, 1.397, 1.439, 1.436, 1.437, 1.438}, <$d_{B-N}$> = 1.424

(d) At the bottom edge (from left to right, in Å ):

    $d_{B-C}$ = {1.747, 1.779, 2.549, 1.747, 1.779, 2.549}, <$d_{B-C}$> = 2.025

    $d_{N-C}$ = {2.708, 2.727, 3.614, 2.708, 2.727, 3.613}, <$d_{N-C}$> = 3.016

    $d_{B-N}$ = {1.449, 1.446, 1.441, 1.449, 1.446, 1.441}, <$d_{B-N}$> = 1.445

(e) At the bottom edge (from left to right, in Å ):

    $d_{B-C}$ = {3.242, 3.306, 3.376, 3.241, 3.305, 3.376}, <$d_{B-C}$> = 3.308

    $d_{N-C}$ = {3.462, 2.877, 2.956, 3.462, 2.876, 2.954}, <$d_{N-C}$> = 3.098

    $d_{B-N}$ = {1.377, 1.432, 1.395, 1.377, 1.432, 1.395}, <$d_{B-N}$> = 1.401

(f) At the bottom edge (from left to right, in Å ):

    $d_{B-C}$ = {1.815, 1.623, 1.563, 1.816, 1.622, 1.563}, <$d_{B-C}$> = 1.667

    $d_{N-C}$ = {2.806, 2.824, 2.868, 2.806, 2.824, 2.868}, <$d_{N-C}$> = 2.833

    $d_{B-N}$ = {1.416, 1.414, 2.749, 1.416, 1.414, 2.749}, <$d_{B-N}$> = 1.860

(g) At the bottom edge (from left to right, in Å ):

$d_{B-C} = \{2.507, 3.162, 3.123, 3.178, 3.230, 1.760\}, \langle d_{B-C}\rangle = 2.827$

$d_{N-C} = \{1.503, 3.070, 3.094, 3.251, 3.203, 3.190\}, \langle d_{N-C}\rangle = 2.885$

$d_{B-N} = \{1.358, 1.412, 1.295, 1.303, 1.388, 1.890\}, \langle d_{B-N}\rangle = 1.441$

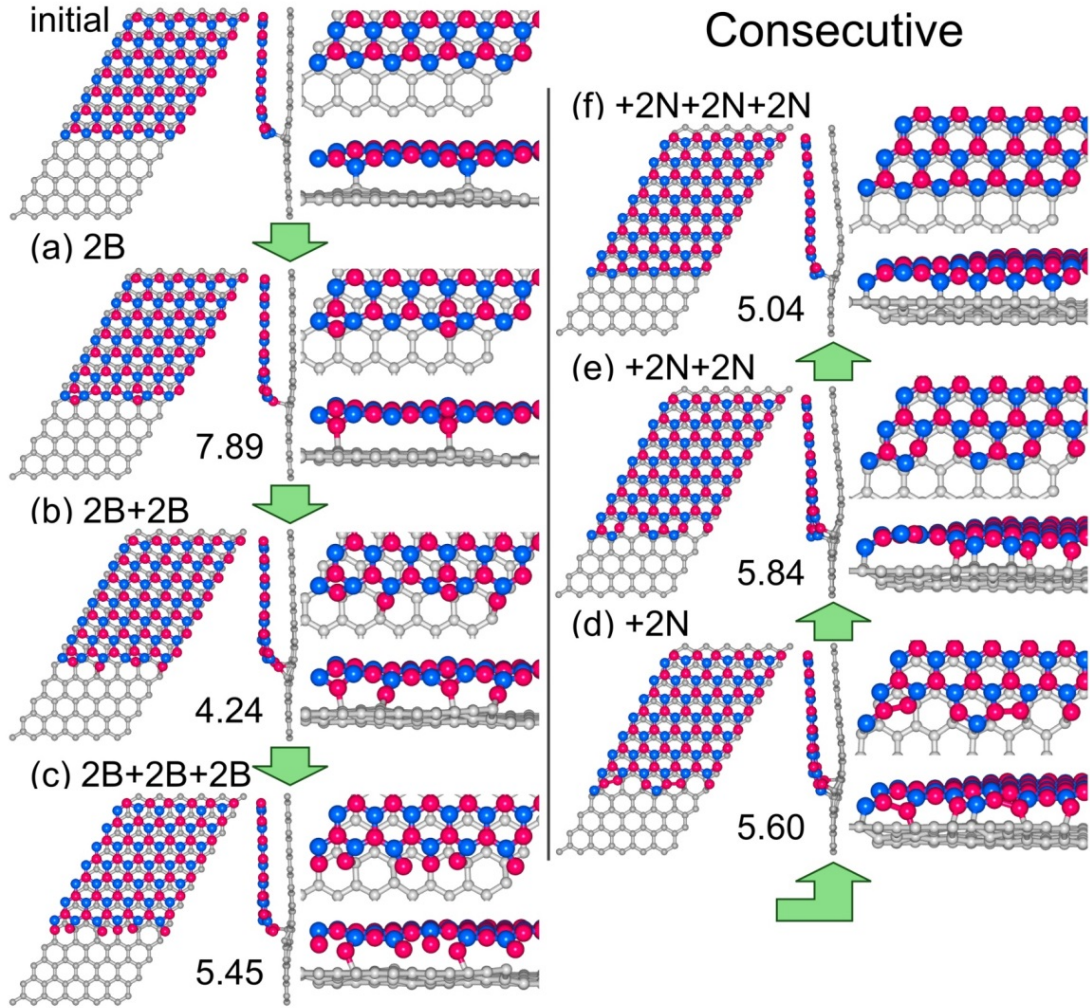

**Fig. S15:** Structural geometries and energies of the consecutive steps in the low-flux (LF) mode of the 1D growth of the h-BN film on the graphene layer. Boron (B) atoms are depicted as red balls, nitrogen (N) atoms in blue, and C atoms in grey color. The process starts by the addition of boron atoms to the initial configuration shown at the top left. (a-c) Three steps (2B, 2B+2B, and 2B+2B+2B) comprising the consecutive boron deposition stage, followed in (d-f) by the subsequent consecutive deposition of nitrogen atoms, resulting in the growth of an added ordered row of hexagons at the growing edge of the h-BN film (see frame (f)). In each of the panels (except the top left) we include the DFT-calculated binding energy (per B or N atom) in eV units, associated with the corresponding stage of h-BN film growth shown in the panel.

Geometrical parameters:

(a) At the bottom edge (from left to right, in Å):

$d_{B-C} = \{1.758, 1.758\}$, $<d_{B-C}> = 1.758$

$d_{N-C} = \{2.818, 2.827, 2.845, 2.818, 2.827, 2.845\}$, $<d_{N-C}> = 2.830$

$d_{B-N} = \{1.444\ 1.444\}$, $<d_{B-N}> = 1.444$

(b) At the bottom edge (from left to right, in Å):

$d_{B-C} = \{1.629, 1.575, 1.629, 1.576\ \}$, $<d_{B-C}> = 1.602$

$d_{N-C} = \{2.829, 2.733, 2.819, 2.829, 2.733, 2.819\}$, $<d_{N-C}> = 2.794$

$d_{B-N} = \{1.462, 1.416, 1.462, 1.416\ \}$, $<d_{B-N}> = 1.439$

(c) At the bottom edge (from left to right, in Å):

$d_{B-C} = \{2.538, 1.681, 1.834, 2.538, 1.681, 1.834\}$, $<d_{B-C}> = 2.018$

$d_{N-C} = \{3.605, 2.657, 2.823, 3.605, 2.657, 2.823\}$, $<d_{N-C}> = 3.028$

$d_{B-N} = \{1.436, 1.435, 1.489, 1.436, 1.435, 1.489\}$, $<d_{B-N}> = 1.453$

(d) At the bottom edge (from left to right, in Å):

$d_{B-C} = \{\ 2.606, 1.757, 1.666, 2.608, 1.764, 1.655\}$, $<d_{B-C}> = 2.009$

$d_{N-C} = \{1.555, 1.559\}$, $<d_{N-C}> = 1.557$

$d_{B-N} = \{1.452, 1.422, 1.442, 1.450, 1.447, 1.446\}$, $<d_{B-N}> = 1.443$

(e) At the bottom edge (from left to right, in Å):

$d_{B-C} = \{2.693, 3.155, 1.681, 2.535, 1.662, 1.662\}$, $<d_{B-C}> = 2.232$

$d_{N-C} = \{1.553, 3.022, 1.555, 1.566\}$, $<d_{N-C}> = 1.924$

$d_{B-N} = \{1.439, 1.300, 1.411, 1.443, 1.412, 1.410\}$, $<d_{B-N}> = 1.402$

(f) At the bottom edge (from left to right, in Å):

$d_{B-C} = \{2.616, 2.608, 2.474, 2.522, 2.525, 2.471\}$, $<d_{B-C}> = 2.536$

$d_{N-C} = \{1.539, 2.568, 1.538, 1.551, 1.558, 1.551\}$, $<d_{N-C}> = 1.717$

$d_{B-N} = \{1.397, 1.397, 1.439, 1.436, 1.437, 1.438\}$, $<d_{B-N}> = 1.424$